\documentclass[11pt]{article} 
\usepackage{epsfig,amssymb} 
\title{\Large\bf Finitely coordinated models for low-temperature phases of 
amorphous systems}  %
\author{Reimer K\"uhn$^1$, Jort van Mourik$^2$, 
Martin Weigt$^3$ and Annette Zippelius$^4$        \\
$^1$Department of Mathematics, King's College London, Strand, 
    London WC2R 2LS, UK\\ 
$^2$NCRG, Aston University, Aston Triangle,  Birmingham B4 7ET, UK\\
$^3$Institute for Scientific Interchange, Viale S. Severo 65,  10133 Torino,
    Italy\\
$^4$Institut f\"ur Theoretische Physik, Universit\"at G\"ottingen\\
    Friedrich-Hund-Platz 1 , 37077 G\"ottingen, Germany\\
}
\date{\today}
\font\tams                   = cmmib10 
\font\kleinhalbcurs          = cmmib10 scaled 833 
\font\bxf                    = cmbx10 
\font\sevenbf                = cmbx7 
\def\vec#1{{\textfont1=\tams\scriptfont1=\kleinhalbcurs 
\textfont0=\bxf\scriptfont0=\sevenbf 
\mathchoice {\hbox{$\displaystyle#1$}}{\hbox{$\textstyle#1$}} 
{\hbox{$\scriptstyle#1$}}{\hbox{$\scriptscriptstyle#1$}}}} 

\def \vr{\vec{r}} 
\def \vR{\vec{R}} 
\def \vp{\vec{p}}  
    
\def \vv{\vec{v}}   \def \tvv{\tilde\vec{v}}
\def \vu{\vec{u}}   \def \tvu{\tilde\vec{u}}

\def \ul{\underline }

  \def \hx{\hat{x}} 
  \def \hX{\hat{X}}

\def \ux{\ul{x}}  \def \uhx{\ul{\hx}}
\def \uX{\ul{X}}  \def \uhX{\ul{\hX}}

\def \hx{\hat x}
\def \hpi{\hat\pi}
\def \hZ{\hat Z}
\def \v0{\vec{0}}

\def \cD{{\cal D}}

\def \cN{{\cal N}}
\def \cO{{\cal O}}
\def \cP{{\cal P}}
\def \cU{{\cal U}}

\def \cT{{\cal T}}

\def\real{{\rm I\!R}} 

\def \rd{\mbox{\rm d}}

\def \nn{\nonumber}
\def \beq{\begin{equation}} 
\def \eeq{\end{equation}} 
\def \bea{\begin{eqnarray}} 
\def \eea{\end{eqnarray}} 
\def \bay{\begin{array}} 
\def \eay{\end{array}}

\newcommand {\np  } {\newpage}

\newcommand {\ov  } {\over   }
\newcommand {\ev  } {\equiv  }

\newcommand {\be  } {\beta   }
\newcommand {\om  } {\omega  }
\newcommand {\de  } {\delta  }
\newcommand {\la  } {\lambda }

\newcommand {\hrh } {\hat\rho  }
\newcommand {\hps} {{\hat{\psi}}}

\newcommand {\cG  } {{\cal G}  }

\newcommand {\ph  } {\varphi   }
\newcommand {\uph} {{\ul{\varphi}}}

\newcommand {\Bra } {\left \langle} \newcommand {\Ket } {\right \rangle}
\newcommand {\lh  } {\left (      } \newcommand {\rh  } {\right )      }
\newcommand {\lv  } {\left [      } \newcommand {\rv  } {\right ]      }
\newcommand {\lc  } {\left \{     } \newcommand {\rc  } {\right \}     }

\newcommand {\scal}[2] {{\left( #1\left |\right . #2\right )    }}
\newcommand {\Tr   } {\mathop{\mbox{\rm Tr}}   }

\renewcommand{\i   } {{\rm i}}
\newcommand {\e    } {{\rm e}}
\newcommand {\bC    } {{\bf C}}
\renewcommand {\d  } {{\rm d}}

\begin{document}  
\maketitle 

\begin{abstract}
We introduce models of heterogeneous systems with finite connectivity defined on
random graphs to capture finite-coordination effects on the low-temperature 
behavior of finite dimensional systems. Our models use a description in terms of 
small deviations of particle coordinates from a set of reference positions, 
particularly appropriate for the description of low-temperature phenomena. A 
Born-von-Karman type expansion with random coefficients is used to model effects 
of frozen heterogeneities. The key quantity appearing in the theoretical description 
is a full distribution of effective single-site potentials which needs to be determined 
self-consistently. If microscopic interactions are harmonic, the effective single-site 
potentials turn out to be harmonic as well, and the distribution of these single-site
potentials is equivalent to a distribution of localization lengths used earlier in the 
description of chemical gels. For structural glasses characterized by frustration and 
anharmonicities in the microscopic interactions, the distribution of single-site
potentials involves anharmonicities of all orders, and both single-well and
double well potentials are observed, the latter with a broad spectrum of barrier
heights. The appearance of glassy phases at low temperatures is marked by the
appearance of asymmetries in the distribution of single-site potentials, as
previously observed for fully connected systems. Double-well potentials with a
broad spectrum of barrier heights and asymmetries would give rise to the well 
known universal glassy low temperature anomalies when quantum effects are taken 
into account.
\end{abstract}
\date
\np
\setcounter{page}{1}
\section{Introduction \label{sec:int}}

Understanding the behaviour of amorphous systems, both at elevated and and at
low temperatures, continues to be a challenging problem. 

Structural glasses, for instance, show intriguing dynamic and thermodynamic
behaviour when they are formed, e.g., by cooling from the liquid state
\cite{Donth01}. A very sharp rise of relaxation times from microscopic to
macroscopically large within a fairly narrow temperature interval entails that
such systems will fall out of (local) equilibrium at a (cooling-rate dependent)
calorimetric glass-transition temperature $T_g$, accompanied by a sharp drop of
the specific heat $c_p(T)$ and thermal expansion coefficient
$\alpha(T)$. Signatures of a dynamic singularity are, however, observed in many
glasses at a temperature $T_c$ {\em larger\/} than that of the calorimetric
transition \cite{Donth01, Goetze91,EdiAngNag96}. Such a dynamic singularity was
shown to emerge within a mode coupling theory (MCT) --- a description of the
dynamics  of supercooled liquids in terms of correlation functions  --- via
non-linear feedback mechanisms and memory-effects. The dynamic singularity
occurs at a temperature where the viscosity is still fairly low; for a detailed
review, see \cite{Goetze91}.
Conversely, if the behaviour of the entropy of the supercooled liquid is
extrapolated to {\em lower\/} temperatures, it would become smaller than the
entropy of the crystal at the Kauzmann temperature $T_K$, which is speculated to
mark the location of a hypothetical true equilibrium glass transition, as it
would be observed in the limit of infinitely slow cooling \cite{Kauzm48}. The
Kauzmann temperature is usually quite close to the temperature $T_0$ marking a
hypothetical divergence of the viscosity when fitted according to a
Vogel-Fulcher-Tammann form, and the two temperatures are often identified (and
ought to be according to classical semi-microscopic \cite{GibDiMar58} and
phenomenological \cite{AdaGib65} approaches to the glass transition).

In a series of seminal papers, Kirkpatrick, Thirumalai, and  Wolynes
\cite{KirkWol87,KirkThir87a,KirkThir87b} pointed out a connection between the
transition to frozen order in certain spin-glass models and the structural glass
transition. The spin-glass models in question exhibit an equilibrium phase
transition to a low-temperature phase with one 1 step replica-symmetry broken
structure of the matrix of spin-glass oder parameters. The transition is
discontinuous in the order parameters, but exhibits no latent heat. Remarkably
there is also a dynamical freezing transition at a temperature $T_d$ that is
different (and higher) than the temperature $T_0$ at which the equilibrium
transition occurs \cite{KirkWol87, KirkThir87a, KirkThir87b}. The phenomenology
is strikingly similar to that found in structural glasses, where the dynamical
transition is associated with the mode coupling transition, while the
equilibrium transition would be associated with a hypothetical equilibrium
transition at the Kauzmann temperature $T_k$, and in both cases, the equilibrium
transition appears to be driven by an entropy crisis. The analogy was further
elucidated and strengthened by studies of {\em spherical\/} version of the
$p$-spin interaction spin-glass model \cite{CriSom92, CriHorSom93}, for which
macroscopic dynamical evolution equations for correlation and response functions
can be formulated in closed form and have in fact the structure of idealized MCT
equations \cite{CriHorSom93}. In this case -- unlike in structural glasses ---
the off-equilibrium dynamics in the non-ergodic low-temperature phase is also
understood in rather great detail \cite{CuglKur93}.

There is, of course, a sense in which the analogy between the physics of
spin-glasses in the $p$-spin class and the physics of structural glasses is
purely formal, as spin-glasses are systems with quenched disorder describing the
interaction of degrees of freedom that are localized at all temperatures,
whereas in structural glasses the disorder is self-induced and  localization
occurs only in the low-temperature phase. The issue of quenched disorder  can
be addressed at least schematically by noting that models with quenched
disorder can, and have been used as bona-fide approximants to describe the
physics of  models with self-induced disorder 
\cite{BouMez94,MarParRit94a,MarParRit94b}, yet the issue of localization
remains. Significant progress has, however, been made towards a  first
principles description of glassy thermodynamics by M\'ezard and Parisi
\cite{MezPar99}, based on an investigation of weakly coupled real replicas of
liquid  systems, elaborating on and extending related ideas of Monasson
\cite{Monas95}, Franz and Parisi \cite{FranPar95a, FranPar95b}. A more recent
development are so-called lattice glasses 
\cite{FrMeRiWeZe01,BiMe02,HaWe03,CiTaCaCo03,RiBiMeMa04,HanWe05}, 
{\it i.e.} lattice gas models with discrete
degrees of freedom on (generalized) Bethe lattices. These models are neither 
locally disordered nor of infinite connectivity, allowing in this way for an 
ordered, crystalline state, and showing again glassy behavior due to
self-induced disorder.

On a quite different note, structural glasses and other amorphous systems were
found to exhibit thermodynamic, dynamic and transport properties at low
temperatures, which are strikingly different from those of their crystalline
counterparts \cite{ZellPohl71}. At temperatures below 1 K, the specific heat of
glasses varies approximately linearly with temperature $T$, the behaviour of the
thermal conductivity is close to quadratic in $T$, whereas in crystals both
quantities are {\em cubic\/} in temperature. The behaviour of acoustic and
dielectric response is also anomalous in this temperature range, when compared
with the corresponding behaviour of crystals \cite{Piche+74}.

At temperatures at around 10 K there is an anomaly of a different kind in the
specific heat which appears as a bump in the ratio $C(T)/T^3$  --- the so-called
Bose peak. The thermal conductivity exhibits a plateau in the Bose peak region,
and begins to rise again as the temperature is further increased.

Below 1 K, the low-temperature anomalies, as the phenomena just described have
become to be referred to, are attributed to the existence of tunnelling systems
which are absent in ideal crystals \cite{And+72,Phil72}. Tunnelling systems are
thought to be realized as double-well configurations in the potential energy
landscape, between which atoms, or groups of atoms can tunnel, even when thermal
energies to overcome the intervening potential energy barriers are not
available. Double-well potentials with a broad range of  asymmetries and barrier
heights can give rise to the  broad spectrum of low-energy tunneling-excitations
responsible for the anomalous thermal properties, whereas their resonant and
relaxational interaction with the phonon bath creates anomalous response
properties. At temperatures (energies) in the Bose peak region, the main
excitations appear to be soft quasi-harmonic. Their origin and precise nature is
much debated (e.g. \cite{Karp+83, Par94, RamBuch98, Schir+98, Grig+03}); in any
case, strong hybridization between these states and the phonon bath  is likely
to be an important feature.

One of the remarkable aspects of glassy low-temperature anomalies is their
considerable degree of universality. Two main contending theorys are available
to explain this fact: Following ideas of Yu and Leggett \cite{YuLegg88},
universality has been understood as a collective effect due to {\em interactions
between\/} quantized excitations, i.e. the tunnelling systems \cite{BurKag96a};
alternatively universality is regarded as a consequence of the irregularities of
the potential energy landscape that are frozen in at the glass transition, {\em
leading to a broad spectrum of\/} quantized low-energy tunneling excitations
\cite{KuHo97, KuUr00}. The potential energy landscape being itself a collective
affair generated through the interactions of constituent particles, one is led
to expect a certain degree of insensitivity to detail of the low-temperature
anomalies. This line of reasoning has been made more explicit in \cite{Kuehn03},
and the key feature responsible for the universality of the  low-temperature
anomalies was demonstrated to be a clear separation of the energy scales frozen
in at the glass transition on the one hand side, and that responsible  for the
low temperature physics on the other hand side. A simple explanation for the
mysterious so-called quantitative universality, according to which not only
power-laws describing the variation of some thermodynamic functions with
temperature, but also prefactors in these power-laws are insensitive to detail,
was also provided \cite{Kuehn03}.

On the other hand, the description of the glassy freezing transition in these
models  is {\em not\/} as one would want to have it for structural glasses, as
the transition to the non-ergodic low-temperature phase is not of the
discontinuous 1-step replica-symmetry breaking (RSB) type as in the $p$-spin models, 
so the description chosen in \cite{KuHo97, KuUr00, Kuehn03} does appear to miss 
essential aspects of glass transition physics. To be fair though, it should be 
mentioned that those models were initially constructed mainly to provide a 
description of low-temperature phenomena. However, being microscopic the question 
as to their high-temperature properties naturally arises, and in this regime one 
might regard them as inadequate in their present form. Conversely, regarding
low-temperature phenomena, it is difficult to conceive how models of the
$p$-spin type would describe the physics of tunnelling systems; in their
spherical mean-field variant in particular \cite{CriSom92, CriHorSom93}, they
would give rise to {\em harmonic\/} effective single-site potentials which could
not support tunnelling systems. It is thus perhaps fair to say that a
description of both, the low-temperature phenomena, and glass-transition physics
within a single unified approach still eludes us.

The main purpose of the present investigation to improve upon the models
investigated in \cite{KuHo97, KuUr00, Kuehn03}, and consider variants in which
interactions are {\em finitely\/} coordinated, keeping them random though, as a
schematic way of modelling a frozen glassy state. Our investigation will exploit
the fact that that the understanding of dilute random systems --- introduced
many years ago by Viana and Bray \cite{ViaBray85} --- has witnessed significant
progress in the last few years, regarding their equilibrium properties 
\cite{Mo98,MezPar01,HaWeBook05}, as well as their dynamic
behaviour \cite{SmerjCugl003, SmerjWeigt003,HanWe05b,Hatchett+05,MoSe05}. 
This progress was also the basis for the previously discussed lattice-glass models, 
which, however, due to their discrete nature do not allow for the description of 
low-temperature excitations.

It is hoped that the additional element of realism introduced by considering
finitely coordinated systems, respecting in this feature an important property
of finite dimensional systems, could help bring those models closer to reality.

The remainder of the paper is organized as follows. In Sect. \ref{sec:Mod} we
introduce our model, and demonstrate how a finitely coordinated random model
arises as a natural candidate to describe the physics of a glassy low
temperature phase. We will look in detail only at a simplified version with
scalar degrees of freedom. We describe the solution of the model using replica
and mean field techniques in Sect. \ref{sec:repl}. However, in Sect. \ref{sec:rs}, 
we evaluate only a replica-symmetric (RS) approximation to the full theory. It
turns out to have a similar complexity as a 1-RSB solution in models with
discrete degrees of freedom. The key quantity appearing in the theoretical
description is a full distribution of effective single-site potentials which
needs to be determined self-consistently. A population-based stochastic
algorithm is used to solve these RS self-consistency equations. The general
set-up is very flexible and can be investigated for a broad spectrums of
assumptions about the nature of the microscopic interactions as well as the
connectivity distributions.

In Sect. \ref{sec:harm} we evaluate the theory for models in which the
microscopic interactions are purely harmonic, in which case the effective
single-site potentials turn out to be harmonic as well; the distribution of the
single-site potentials is then equivalent to a distribution of localization
lengths used earlier in the description of systems of such type \cite{Goldb+96,
ZipGol98, Brod+02,Mao+05,Mao+06}, which may be thought of as
systems in their gel-phase. For structural glasses characterized by frustration
and anharmonicities in the microscopic interactions, investigated in Sect
\ref{sec:anharm}, the distribution of single-site potentials involves
anharmonicities of all orders, and both single-well and double well potentials
are observed, the latter with a broad spectrum of barrier heights. Although
initially designed as models for the description of the low-temperature phase,
it makes perfect sense to consider them also at high temperatures, where the
microscopic symmetries of the interaction energy are unbroken. The appearance of
glassy phases, as the temperature is lowered is marked by the appearance of
asymmetries in the distribution of single-site potentials, as previously
observed for fully connected systems. We investigate the phase diagram for a 
model system of this kind and compute thermodynamic function across the 
transition. As in fully connected systems, the transition is found to be 
continuous in our anharmonic model, and in this aspect the model we are 
considering here remains deficient. We discuss possible remedies of this 
deficiency in a concluding section.  

Details of some of the calculations were relegated to appendices. The paper
closes with a summary and an outlook on future interesting lines of research.

\section{The Model}						\label{sec:Mod}

We consider a many-particle system described by the Hamiltonian:
\beq
H=\sum_i{\vp_i^2\ov 2m}+U_{\rm int}(\{\vr_i\})
\eeq
with an interaction energy taking the form
\beq
U_{\rm int}=\sum_{(i,j)}\phi(\vr_i-\vr_j)+\sum_i V(\vr_i)
\label{Uint0}
\eeq
Our initial goal is to attempt a microscopic, albeit schematic description of
the physics of an amorphous random solid. We proceed as follows. To begin with,
we decompose each particle coordinate $\vr_i$ into a fixed reference coordinate
$\vR_i$ and a relative coordinate $\vu_i$ characterizing the (small) excursion
from their respective reference position. Supposing that the initial pair
potential has an effectively finite range, the interaction potential can then be
formulated in terms of local connectivity variables and a (random) potential
depending only on the relative coordinates.

The connectivity is characterized by a matrix ${\bf C}$ with elements $c_{ij}$,
such that $c_{ij}=1$ for close particles, for which $|\vR_i-\vR_j| \le r_c$,
with $r_c$ denoting the interaction range of the initial pair potential $\phi$,
and $c_{ij}=0$ if $|\vr_i-\vr_j| \simeq  |\vR_i-\vR_j| \gg r_c$ and so
$\phi(\vr_i-\vr_j)\simeq 0$. The interaction potential (\ref{Uint0}) may then
be rewritten as
\beq 
U_{\rm int}=\sum_{(i,j)}c_{ij}\phi_{ij}(\vu_i-\vu_j)+\sum_i V_i(\vu_i)
\label{Uintv1}
\eeq 
Here the symbol $(i,j)$ is used to denote pairs of sites; non-zero pair potentials 
$\phi_{ij}$ and single-site potentials $V_i$ may vary from pair to pair and site 
to site, thus introducing residual disorder (arising, e.g., from  interactions 
between different species, random nonzero $\vR_i-\vR_j$, etc., in the case of 
pair potentials, and the coupling of different species to external fields in 
case of the single-site potentials); we will specify  $\phi$ and $V$ only later. 
One may think of the $\phi_{ij}(\vu_i-\vu_j)$ as of a  Born-von Karman expansion 
of the interaction potential about the reference positions, carried to all orders.

In case one is interested mainly in low-temperature phenomena, one would expect
that some low order expansion of the $\phi_{ij}$ should be sufficient to
capture the essential physics.

As a crucial last ingredient, which renders the model solvable, we consider
connectivity matrices $\bC=\{c_{ij}\}$ which are not created as a consequence of
an underlying geometrical arrangement of the interacting particles as described
above, but rather we take them to be random.
We assume that the local coordination number $L_i$ of site $i$ is distributed
according to a distribution $P_C(L)$, with {\it finite\/} average connectivity
$C= \sum_L L P_C(L) =  {1\ov N}\sum_iL_i$, and that the probability that two 
sites $i\neq j$ with coordination numbers $L_i,~L_j$ are connected (i.e. that 
$c_{ij}=c_{ji}=1$) is proportional to $L_iL_j$ (no correlations).
The adjacency matrices $\bC$ in the ensemble determined by $P_C(L)$ then
have the following distribution
\beq
P_c({\bC|\{L_i\}})={1\ov \cN} \prod_{(i,j)} p_C(c_{ij}) ~\de_{c_{ij},c_{ji}}
\prod_i \de_{L_i,\sum_{j(\neq i)}c_{ij}}
\label{P_ofC}
\eeq
where $\cN$ is a normalization constant. It iturns out that the distribution 
(\ref{P_ofC}) is {\em independent\/} of the conditioning $\{L_i\}$ in the large $N$ 
limit, for {\em any typical\/} realization of the  set $\{L_i\}$ compatible with 
a given distribution 
\beq
P_C(L) = \frac{1}{N} \sum_i \de_{L,L_i}
\label{Pc_ofL}
\eeq
of coordination numbers. {\it A priori} single bond probabilities
\beq
p_C(c_{ij}) = {C\ov N}\de_{c_{ij},1}+\left(1-{C\ov N}\right)\de_{c_{ij},0}\ .
\eeq
compatible with the average connectivity $C$ are also included in (\ref{P_ofC}), as
a matter of convenience rather than necessity, as constraints coming from the full 
distribution of coordination numbers are also encoded in (\ref{P_ofC}). Other, 
equivalent formulations are possible as noted in Appendix \ref{app:replica_det}. 
Note that for $P_C(L)= {C^L\ov L!}\exp(-C)$ the connectivity matrix is that of 
an Erd\"os-Renyi random graph. 

The residual randomness in the pair and single-site potentials is described by  
normalized probability densities $P_\phi[\phi]$ and $P_V[V]$. 

In its present formulation, the model can be solved within replica mean-field
theory. The nature of the random pair and single-site potentials as well as the
distribution of connectivities can be left open for the time being. Specifying them 
in different ways one has access to describing a variety of different physical systems. 

\section{Replica Analysis}					 \label{sec:repl}

The generator of all the relevant physical quantities is the free energy, 
$-N\be f(\be)=\ln(Z_N)$, and using the standard replica trick
$\Bra\log Z \Ket=\lim_{n\to0} n^{-1} \log\Bra Z^n\Ket$, we need to
calculate the average of the $n$-fold replicated partition function
over the bond-disorder, i.e. the $\bC$, and the $\phi_{ij}$:
\beq
\Bra Z_N^n\Ket_{\bC,\phi}= \Bra \int \prod_{i,a}d \vu_i^a\exp\lc -\lh\be
\sum_{(i,j),a}c_{ij}\phi_{ij}(\vu^a_i-\vu^a_j)+\sum_{i,a} V_i(\vu^a_i)
\rh\rc\Ket_{\bC,\phi}
\label{reppartfunc}
\eeq
Free energies are manifestly self-averaging w.r.t. any form of (finite-dimensional) 
site disorder, which is why taking a $V$-average is not required here.
It is useful to introduce the short-hand notation $\tvu=(\vu^1,\dots,\vu^n)$ for a 
{\it replicated vector}, and the {\it replica} interaction (bond) and single-site 
weights:
\beq
\cU_b(\tvu,\tvv,\phi)\ev\exp\lh-\be\sum_a\phi(\vu^a-\vv^a)\rh
~,\quad
\cU_s(\tvu     ,V   )\ev\exp\lh-\be\sum_a V  (\vu^a      )\rh\ .
\label{rpot}
\eeq
The line of reasoning involved in performing the average over connectivity matrices
for non-Poissonian connectivity distribution closely follows \cite{WongSherr87}. 
Constraints on the local connectivities in the $\bC$-average are enforced via the 
identity
\beq
\de_{K_i,L_i}=\oint {dz_i\ov 2\pi\i z_i} z_i^{(K_i-L_i)}
\label{Kron_delt}
\eeq
It is found that the average of the replicated partition function can be written as
a functional integral over the {\it `replica density'}
\beq
\rho(\tvu)\ev{1\ov N}\sum_i z_i~\de(\tvu-\tvu_i)\ .
\label{repdens}
\eeq
Enforcing its  definition in terms of conjugate variables $\hrh(\tvu)$
we obtain after some standard steps (for details see appendix A):
\beq
\Bra Z_N^n\Ket_{\bC,\phi}=\frac{1}{\cN}\int\cD\rho~\cD\hrh~\exp\lc N\left(
{C\ov2}\Big(G_b[\rho]-1\Big)-G_m[\rho,\hrh]+G_s[\hrh]\right)\rc\ ,
\label{ZG}
\eeq
where
\bea
G_b[\rho     ~]&=&\Bra
   \int d\tvu~d\tvv~\rho(\tvu)    ~\cU_b(\tvu,\tvv,\phi)~\rho(\tvv)\Ket_\phi
\nn\\
G_m[\rho,\hrh~]&=&    
   \int d\tvu~d\tvv~\hrh(\tvu)~                \rho(\tvu) 
\label{defG}\\
G_s[     \hrh~]&=&\sum_L P_C(L)~\Bra\ln
   \int d\tvu~\hrh^L(\tvu)~\cU_s(\tvu,        V)          \Ket_V
\nn
\eea
The path integral is dominated by its saddle point, and functional variation 
with respect to $\rho(\tvu)$, and $\hrh(\tvu)$ leads to the stationarity
conditions:
\bea
\hrh(\tvu)&=&C \Bra\int d\tvv~\cU_b(\tvu,\tvv,\phi)~\rho(\tvv)\Ket_\phi
\label{fullFPE1}\\
\rho(\tvu)&=&\sum_{L}P_C(L)L\Bra{
                                 \hrh^{L-1}(\tvu)~\cU_s(\tvu,V)\ov
                      \int d\tvv~\hrh^L      (\tvv)~\cU_s(\tvv,V)   }\Ket_V
\label{fullFPE2}
\eea
%
\subsection{Replica Symmetry 					\label{sec:rs}}

In order to be able to take the limit $n\to0$ in (\ref{ZG})-(\ref{fullFPE2}) as 
required for the replica method, we need to make an {\em ansatz} for the replicated 
density $\rho$ in (\ref{repdens}) and its conjugate $\hrh$, which assumes certain 
symmetry properties under permutation of the replicas. Within the present paper 
we shall only explicitly deal with a replica symmetric (RS) ansatz. Variants which 
would break the symmetry between replicas are fairly straightforward to write down 
(see Appendix \ref{app:replica_det}), but for the model class studied in the present 
paper they turn out to be extremely complex and difficult to handle numerically. The 
reader will appreciate this statement once the RS theory has been developed.

The RS ansatz describes a situation with unbroken replica symmetry, and without
loss of generality both $\rho(\tvu)$ and its conjugate $\hrh(\tvu)$ can be
written as (functional) superpositions of products of single replica functions
of the form
\bea
\rho(\tvu) &=&  \int\cD\psi~\pi [\psi]~
\prod_a{\exp\lh-\be\psi(\vu^a)\rh\ov Z[\psi]}
\label{rs1} \\
\hrh(\tvu) &=&C \int\cD\hps~\hpi[\hps]~
\prod_a{\exp\lh-\be\hps(\vu^a)\rh\ov Z[\hps]}\ .
\label{rs2}
\eea
Here we use the convention
\beq
Z  [f]  \ev  \int \d \vu~\exp\Big(-\be f(\vu) \Big)\ ,
\label{Zf}
\eeq
i.e. the single replica functions in (\ref{rs1}) and (\ref{rs2}) are taken to have 
Gibbsian form,  with functions $\psi$ and $\hps$ denoting  single-replica potentials that 
generate the Gibbs distributions in question.
Note that $\cD\psi$ and $\cD\hat\psi$ are suitable measures in function space. 
With the prefactor $C$ as chosen in (\ref{rs2}), it turns out that,  both $\pi$ and 
$\hpi$ are normalized probability density functions (pdfs) over function space 
(see Appendix A).
In terms of these specifications, the replicated partition function (\ref{ZG})
may be re-expressed as a functional integral over $\pi$  and $\hpi$ in a way that 
allows to isolate the leading order in $n$ (i.e. $\cO(n)$  in the limit $n\to 0$). 
One obtains an expression of the form
\bea
\Bra Z_N^n\Ket_{\bC,\phi} & \sim & \int\cD\pi~\cD\hpi
\exp\lh nN\lc{C\ov2} \cG_b[\pi]-C\cG_m[\pi,\hpi]+\sum_L P_{C}(L)~\cG_{s,L}[\hpi]
          \rc\rh\ ,
\label{Zfinal}
\eea
On introducing
\beq
Z_2[f,g,h]\ev\int \d \vu~\d\vv~\exp\Big(-\be(f(\vu)+g(\vv)+h(\vu,\vv))\Big)\ ,
\label{Z2}
\eeq
and short hands of the form $\cD\pi[\psi]=\cD\psi~\pi[\psi]$  and 
$\{\cD\hpi\}_L\ev \prod_\ell^L\cD\hpi[\hps_\ell]$,  one can express the 
three functionals appearing in (\ref{Zfinal}) as
\bea
\cG_b[\pi     ]& \simeq & \int\cD\pi[\psi_1]~\cD\pi[\psi_2]~
\Bra\ln\lh{Z_2[\psi_1,\psi_2,\phi]\ov Z[\psi_1]Z[\psi_2]} \rh\Ket_\phi~,
\label{cGb} \\
\cG_m[\pi,\hpi]& \simeq & \int\cD\pi[\psi ]~\cD\hpi[\hps ]~
      \ln\lh{Z[\psi+\hat\psi] \ov Z[\psi] Z[\hps]} \rh~,
\label{cGm} \\ 
\cG_{s,L}[\hpi]& \simeq & \int\{\cD \hpi\}_L~
\Bra\ln\lh {Z\Big[\sum_{\ell=1}^L \hps_\ell+V\Big]\ov
\prod_{\ell=1}^L Z[\hps_\ell]} \rh\Ket_V~.
\label{cGs}
\eea
All $L$ (coordination number) summations are over non-negative integers
(and the convention is used that an empty product evaluates to unity).

Note that one can easily identify a {\em bond} (i.e. pair-interaction) term
$\cG_b$, a mixture term $\cG_m$, and coordination number $L$ dependent single
{\em site} terms $\cG_{s,L}$\/.

The precise form of the various functions and parameters is obtained from the
stationarity condition with respect to variations of $\pi$, which involves
solving stationarity conditions w.r.t. the conjugate functional and $\hpi$ as
well (Lagrange multipliers $\la,~\hat\la$ are introduced to take care of the
normalization constraint on $\pi$ and $\hpi$ in the variational procedure).
We finally obtain (for details see e.g. \cite{Mo98,HaWeBook05}) the following 
coupled set of integral equations for $\hpi$ and 
$\pi$:
\bea
\hpi[\hps] &= &\int\cD\pi[\psi]~\Bra\de\lv\hps-\hat\Psi[\psi,\phi]\rv\Ket_\phi
\label{SP_hpi}\\
\pi[\psi] &=& \sum_L{L\ov C}~P_{C}(L) \int \{\cD\hpi\}_{L-1}~
\Bra\de\lv \psi-\Psi[\{\hps _\ell\}_{L-1},V]\rv\Ket_V~,
\label{SP_pi}
\eea
in which the functions $\Psi[\{\hps_\ell\}_L,V]$ and $\hat\Psi[\psi,\phi]$ are 
defined as
\bea
\Psi[\{\hps_\ell\}_L,V] & = & \sum_\ell^L\hat\psi_\ell+V\ ,
\label{Psi}\\
\hat\Psi[\psi,\phi] & = & -\be^{-1}\ln Z_{\psi,\phi}\ ,
\label{PsiH}
\eea
with 
\beq
Z_{\psi,\phi}(\vv)\ev\int d\vu~\exp\Big(-\be\psi(\vu)-\be\phi(\vu-\vv)\Big)\ .
\label{Zpsiphi}
\eeq
\newline
To summarize, we see that $\hpi[\hps]$ can be obtained from sampling from 
$P_\phi[\phi]$ and $\pi[\psi]$, while $\pi[\psi]$ can be obtained from sampling 
from ${L\ov C}P_C(L)$, $P_V[V]$, and $\hpi[\hps]$. This provides a strategy of
solving the fixed point equations via a population dynamics based algorithm.

In the saddle point the expression for the free energy (see Appendix \ref{app:thd_func})
simplifies to
\bea
\be f(\be) &=& {C\ov 2}\int \cD\pi[\psi_1]\cD \pi[\psi_2]~
\Bra{\ov}\ln Z_2[\psi_1,\psi_2,\phi]\Ket_\phi  \nn \\
& & ~~~~~ -\sum_L P_{C}(L) \int \{\cD \hpi\}_L ~ 
\Bra{\ov}\ln Z[\Psi[\{\hps\}_L,V]]\Ket_V \ .
\label{RSFreeEn}
\eea
The internal energy is given by:
\bea
E(\be) &=& {C\ov 2}\int \cD\pi[\psi_1]\cD\pi[\psi_2]
\Bra\Bra \phi(\vu-\vv){\ov}\Ket_{b,2}\Ket_\phi \nn \\
& & ~~~~~ +\sum_L P_{C}(L) \int  \{\cD \hpi\}_L ~ 
\Bra\Bra V(\vu){\ov}\Ket_{s,L}\Ket_V\ ,
\label{RSIntEn}
\eea
where $\Bra\cdot\Ket_{b,2}$ denotes the Gibbs average corresponding to the bond
energy
\beq
\hat\Psi_b[\psi_1,\psi_2,\phi](\vu,\vv)=\psi_1(\vu)+\psi_2(\vv)+\phi(\vu-\vv)~,
\eeq
while $\Bra\cdot\Ket_{s,L}$ is the Gibbs average corresponding to the
single-site potential $\Psi[\{\hps_\ell\}_L,V]$.
Note that contributions to the internal energy arising from temperature-dependences of 
the distributions $\pi$ and $\hat\pi$ vanish owing to the stationarity condition on 
$f(\be)$ w.r.t $\pi$ and $\hat\pi$.

Thermal averages of arbitrary single-site observables are evaluated as Gibbs 
averages in the ensemble (\ref{Psi}) of effective single-site potentials $V_{\rm eff}= 
\sum_\ell^L\hat\psi_\ell+V$, which are distributed according to
\beq
P[V_{\rm eff}] = \sum_L ~P_{C}(L) \int \{\cD\hpi\}_{L}~
\Bra\de\lv V_{\rm eff} -\Psi[\{\hps _\ell\}_{L},V]\rv\Ket_V~,
\label{PVeff}
\eeq
i.e. for an arbitrary observable of the form $A = \frac{1}{N} \sum_i A(\vu_i)$ we 
have that its thermal average is given by
\bea
\Bra A \Ket &=& \int \cD P[V_{\rm eff}]~\frac{\int \d \vu~A(\vu)~
\exp\big(-\beta V_{\rm eff}(\vu)\big)}{Z[V_{\rm eff}]}\nn\\
            &=& \sum_L ~P_{C}(L) \int \{\cD\hpi\}_{L}~
\Bra \frac{\int \d \vu~A(\vu)~\exp\big(-\beta \Psi[\{\hps\}_L,V](\vu)\big)}
{Z[\Psi[\{\hps\}_L,V]]}\Ket_V
\eea

Note that for Poissonian random graphs, the ensemble of single-replica potentials
$\psi$ is {\em equivalent\/} to the ensemble of effective single-site potentials
$V_{\rm eff}$. This follows by comparison of (\ref{SP_pi}) and (\ref{PVeff}) on 
noting that $\frac{L}{C}P_C(L) = P_C(L-1)$  for Poisson distributions. This implies
in particular that the equivalence between the ensembles of single-replica potentials 
and effective single-site potentials {\em is lost\/} for systems with connectivity 
distributions other than Poissonian.

We note that an alternative way of obtaining the replica symmetric theory is via
the Bethe-Peierls iterative method for obtaining free energies on (locally) tree-like
structures \cite{MezPar01}. Appendix \ref{app:BPMethod} gives a sketch of the reasoning 
for the present model class.

This concludes the general framework.

\subsection{Orthogonal Function Representation}

It should be clear that instead of working with the representation-free 
full functional set-up described above, we could have chosen to represent
the single replica potentials $\psi$ and $\hat\psi$ in terms of their 
expansions using a suitable complete set of basis functions $\{\ph_\mu\}$:
\beq
\psi(\vu)=\sum_\mu x_\mu \ph_\mu(\vu) \ev \ux \cdot\uph(\vu),
\eeq
and analogously $\hat\psi(\vu)=\underline{\hx_\mu}\cdot\uph(\vu)$. Integrals over the
function space are then replaced by multiple integrals over expansion
coefficients. It is convenient to choose the basis functions $\{\ph_\mu\}$
to be orthonormal with respect to a scalar product,
\beq
\scal{\ph_\mu}{\ph_\nu}=\de_{\mu,\nu}~,
\eeq
such that any function $f$ spanned by the $\{\ph_\mu\}$ has an expansion
\beq
f(\vu)=\ul{f}\cdot\uph(\vu),\hspace*{1cm} f_\mu=\scal{f}{\ph_\mu}.
\eeq
Instead of (\ref{rs1}) and (\ref{rs2}) one would write
\bea
\rho(\vu) &=& \ \ \ \int d\pi(\ux)~\prod_a 
\frac{\exp[-\be \ux \cdot\uph(\vu^a)]}{  Z( \ux)}
\label{rs1_n} \\
\hrh(\vu) &=& C      \int d\hpi(\uhx)~\prod_a 
\frac{\exp[-\be \uhx\cdot\uph(\vu^a)]}{\hZ(\uhx)}
\label{rs2_n}
\eea
for the replica density and its conjugate, with $Z(\ux)$ and $\hZ(\uhx)$ fixed
by a normalization condition. Shorthands such as $d\pi(\ux) \ev
d\ux~\pi(\ux)$ are being used in analogy to before.

Repeating the above line of reasoning  one would obtain
\bea
\hpi(\uhx) &= & \int d \pi(\ux)~\Bra \de\lh\hat{\ux}-\uhX(\ux,\phi)\rh \Ket_\phi\\
\label{SP_hpi_n}
\pi(\ux) &=& \sum_L{L\ov C}~P_{C}(L) \int \prod_\ell^{L\!-\!1}
d \hpi(\uhx_\ell) ~ \Bra \de\Big( \ux - \uX_{L-1}(\{\uhx_\ell\})\Big ) \Ket_V
\label{SP_pi_n}
\eea
with
\bea
\uX_{L-1}(\{\uhx_\ell\}) &=& \scal{\uph}{V} + \sum_\ell^{L-1}\uhx_\ell \ ,\\
\label{uXLm1}
\uhX(\ux,\phi)&=& - \be^{-1}\scal{\uph}{\ln Z_{\ux,\phi}}~,
\label{uhXn}
\eea
and
\beq
Z_{\ux,\phi}(v) =\int d\vu~\exp\lh-\be\ux\cdot\uph(\vu)- \be\phi(\vu-\vv)\rh
\label{Z_om_x_n}
\eeq
instead of (\ref{SP_hpi})-(\ref{Zpsiphi}). The distribution (\ref{PVeff}) of
effective single site potentials translates in an obvious way to a corresponding
distribution of their expansion coefficients in an orthogonal function representation,
given by
\beq
P(\ux_{\rm eff}) = \sum_L ~P_{C}(L) \int \prod_\ell^{L}
d \hpi(\uhx_\ell) ~ \Bra \de\Big( \ux_{\rm eff} - \uX_{L}(\{\uhx_\ell\})\Big)  \Ket_V\ .
\eeq
An obvious advantage of explicit
representations of this form is that one may use it to formulate natural
approximation schemes by truncating the expansions at some suitable finite
order. A corresponding disadvantage would be that such truncations can be
expected to be efficient only, if the system of basis functions is well adapted
to the problem being studied.

\section{Harmonic Couplings }				\label{sec:harm}
%
The first model that we investigate is a random graph of harmonically 
coupled {\em scalar\/} degrees of freedom with a distribution of the 
coupling strengths. We assume that there is no single site potential, so
that the system is translationally invariant. The interaction energy 
between connected vertices $(i,j)$ of the graph is given by
\beq
\phi_{ij}(u_i-u_j)={k_{ij}\ov2}(u_i-u_j)^2
\eeq
Models of this type have been used to describe gels; typically, uniform
distributions of cross-link strengths $P(k_{ij})=\delta(k_{ij}-k)$ where 
chosen in that context \cite{Goldb+96, ZipGol98, Brod+02}. Alternatively 
models with random harmonic couplings on regular lattices \cite{Schir+98} 
have been looked at in connection with Bose-peak phenomena. Note that 
\cite{Brod+02}, which --- unlike our present modelling --- provides a 
description preserving information about {\em spatial\/} structure, has 
recently been used to provide a microscopic underpinning of a phenomenological 
model of an elasticity theory for soft random solids based on randomly varying
elastic constants \cite{Mao+06}. In the present investigation, we combine 
elements of random {\em structure\/} with random values for existing 
couplings.

It turns out that the case of harmonically coupled degrees of freedom provides
the only example where the population dynamics described above is closed
within a parameter space of {\em finite\/} dimensionality. In all other cases,
finite dimensional parameter spaces can only give variational approximations
to a full solution.

It should be noted, however, that harmonically coupled systems have unique
ground states and therefore miss essential elements of glassy low-temperature
physics. E.g. they are unable to support two-level tunneling systems in their
potential energy landscape.

On the positive side, one would expect the replica symmetric approximation to be
exact for such systems.

Since the local variables can take any real value, we take the modified
Hermite polynomials $\ph_\mu(u)\ev{1\ov\sqrt{h_\mu}} H_\mu(u)$, with
$h_\mu=\sqrt\pi 2^\mu\mu!$ as the appropriate set of basis functions, in which
the $H_\mu(u)$ are Hermite polynomials satisfying the recursion relation  
\beq
H_{\mu+1}(u)=2u~H_\mu(u)-2\mu~H_{\mu-1}(u)\ ,\hspace*{7mm}\mu=0,1,2..,\infty,
\hspace*{7mm} H_0(u)=1\ . 
\eeq
The polynomials $\ph_\mu$ are orthonormal with respect to the scalar product: 
\beq
\scal{f}{g}\ev\int {du}~\exp(-u^2)~f(u)~g(u) 
\eeq
In the present case the family of orthogonal polynomials can be truncated after
$\ph_2$, to leave the following set of basis functions:
\beq
\{\ph_\mu(u)\}=\left\{{1\ov \sqrt{h_0}}~,~ {2 u\ov \sqrt{h_1}}~
,~{4 u^2-2\ov \sqrt{h_2}}\right\}\ .
\eeq
In the fixed point equations (\ref{SP_hpi_n}-\ref{uhXn}), the integral in
(\ref{Z_om_x_n}), for $\phi(u-v)= \frac{k}{2}(u-v)^2$, is Gaussian and can be 
done analytically, such that the $\hX_\mu$ are determined as:
\beq
\hX_0=x_0-{1\ov \be}\ln \sqrt{2\pi\ov \be k a}
-{\sqrt{h_0}\ov k a}\lh{2 x_1^2\ov h_1}+{16 x_2^2\ov h_2}\rh\ ,\ \ \
\hX_1= {x_1 \ov a}\ ,\ \ \
\hX_2= {x_2 \ov a}\ ,
\label{hXn-harm}
\eeq
where 
$$
a \ev 1+{8 x_2\ov\sqrt{h_2}~k}\ ,
$$
thus leaving us with algebraic updates for the population dynamics only.

We observe that a positive $x_2$ generates a positive $\hX_2$, thus
guaranteeing stability starting from distributions $\pi(\ux)$ and
$\hpi(\uhx)$ which are non-zero for $x_2,\hx_2\in\real^+$ only.
We also note that the evolution of the $x_0$ and $\hat x_0$ distribution
does not couple back into the distribution of non-constant contribution to the
$\psi(u)$ and $\hat\psi(u)$ respectively. Finally, it is interesting to observe
that the distribution of $x_1$, $x_2$, $\hx_1$ and $\hx_2$, are temperature
independent for the harmonic system, while the $x_0$ and $\hx_0$ distributions
exhibit a non-trivial temperature dependence.

The results are presented in the following figures. Fig \ref{Fig:pi_of_x2} shows the
distribution of the expansion coefficient $x_2$ of the single-replica potential 
in an expansion in modified Hermite polynomials as explained above, for a Poissonian
random graph. As mentioned before, this is in the present case equivalent to the 
distribution of the expansion coefficient $x_{{\rm eff},2}$ of the (purely harmonic) 
effective single-site potential. The expansion coefficient $x_{{\rm eff},2}$ is in 
turn related to the localization length $\xi$ of the particles via
\beq
\xi^{-2} = \beta ~\frac{8 x_{{\rm eff},2}}{\sqrt h_2}\ .
\eeq
The multiple peak structure in the distribution of localization lengths is notable,
and related to the fact that localization lengths of different particles are to first
approximation determined by their coordination --- the clearly discernible 7 peaks in
the left panel of Fig \ref{Fig:pi_of_x2} corresponding to coordinations $L=1$ (dangling bond)
to $L=7$. These results have in fact been obtained earlier via different routes 
\cite{Brod+02}. A zoom into the low $x_2$ (large $\xi$) region reveals significant
substructure which shows that correlation lengths of particles depend in fact also 
on coordinations of their neighbours (and on coordinations of next neighbours and so on).

\begin{figure}[!ht]
$$
\epsfig{file=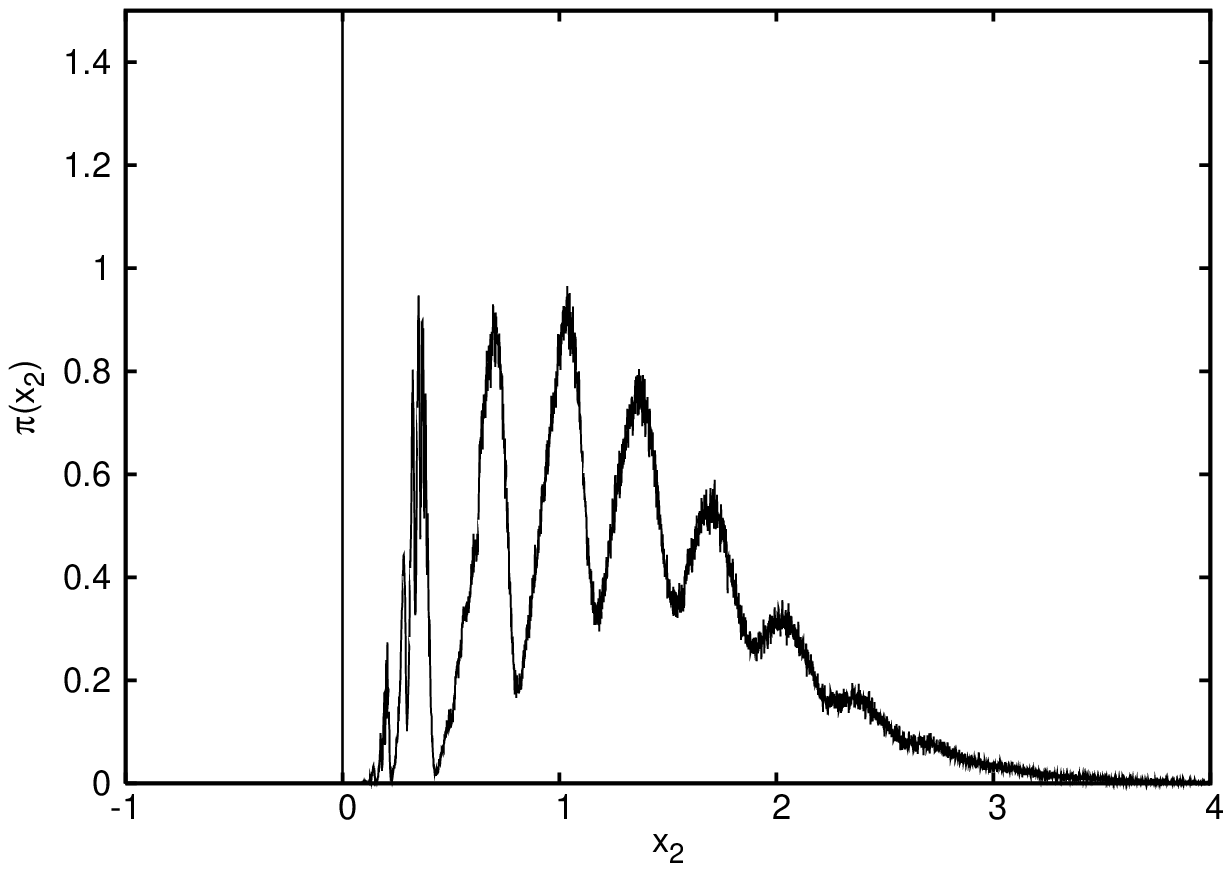,width=0.475\textwidth}
\hfill
\epsfig{file=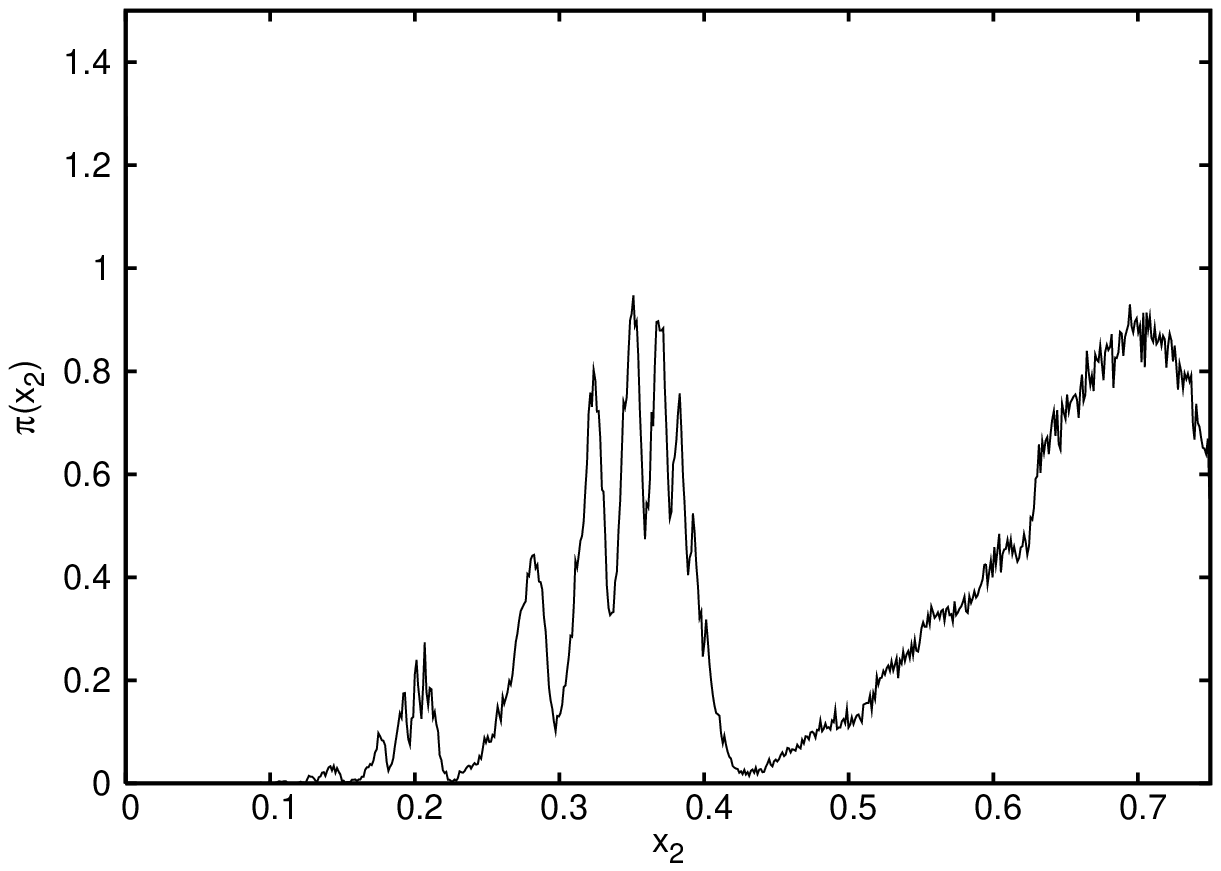,width=0.475\textwidth}
$$
\caption{Distribution $\pi(x_2)$ for a harmonically coupled system with homogeneous
coupling strengths $k_{ij}\ev 1$ on a Poissonian random graph of average connectivity
$C=4$; the inverse temperature is $\beta=1$. Left: full distribution; right: zoom into the 
low $x_2$ region. The $\delta$-peak at $x_2=0$ in the left panel is not drawn to scale; 
its total weight corresponds to the fraction of vertices {\em not\/} in the percolating 
cluster.} \label{Fig:pi_of_x2}
\end{figure}

\begin{figure}[!hb]
$$
\epsfig{file=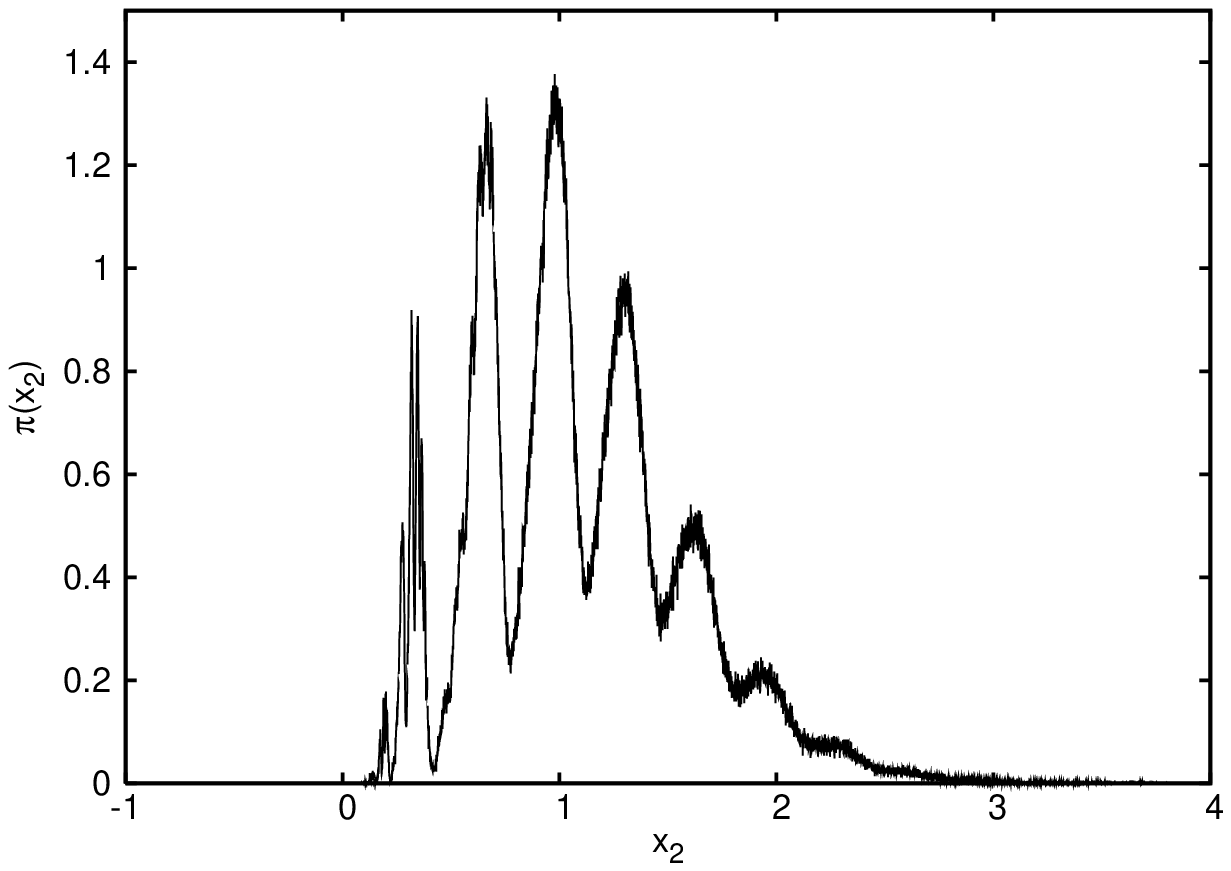,width=0.475\textwidth}
\hfill
\epsfig{file=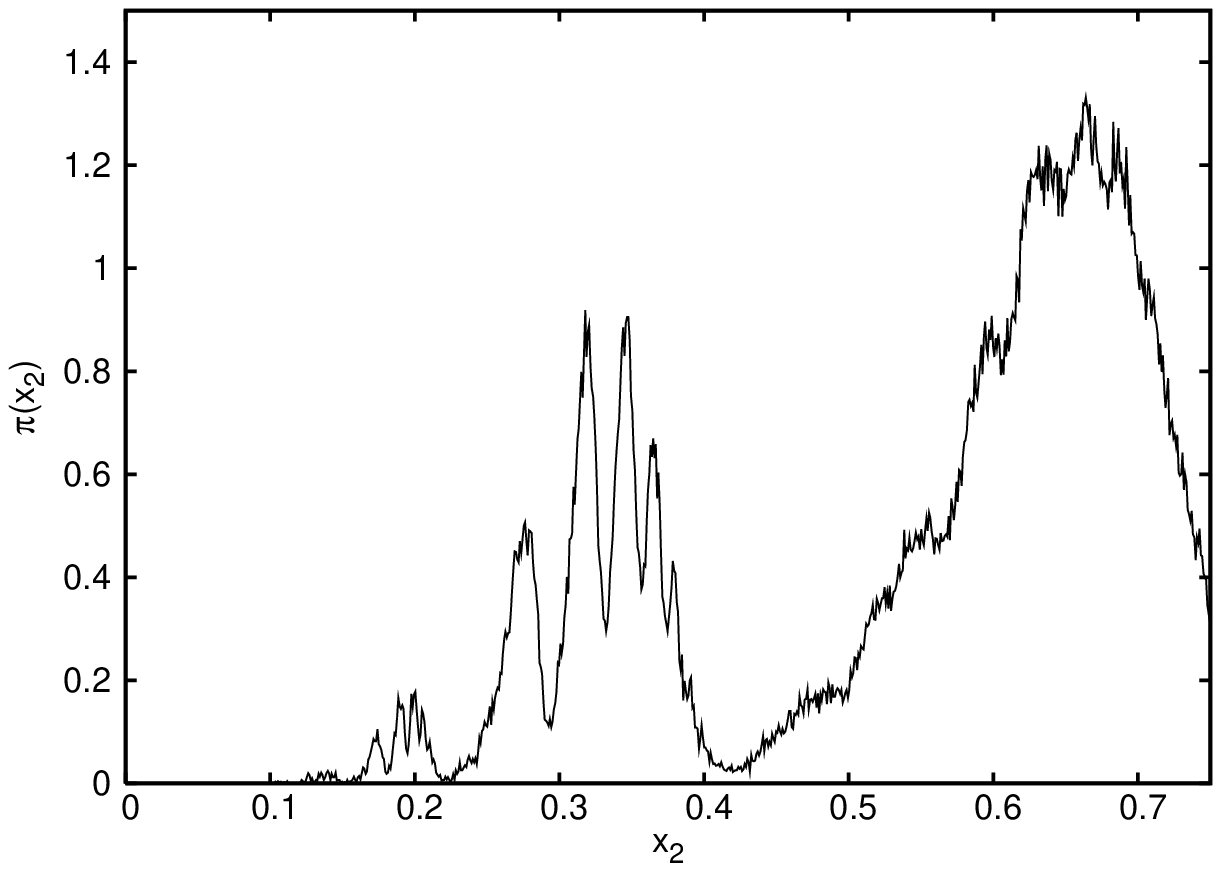,width=0.475\textwidth}
$$
\caption{Distribution $\pi(x_2)$ for a harmonically coupled system with homogeneous
coupling strengths $k_{ij}\ev 1$ on a non-Poissonian random graph of average connectivity
$C=4$ as described in the text; the inverse temperature is $\beta=1$. Left: full 
distribution; right: zoom into the low $x_2$ region.} \label{Fig:pi_of_x2_Lg2}
\end{figure}

In Fig \ref{Fig:pi_of_x2_Lg2} we show analogous results for a graph with non-Poissonian 
degree distribution, assuming connectivities of the form $L=2+L'$ with $L'$ a 
Poissonian of mean 2. In effect, therefore, we do again have mean connectivity $C=4$.
However, there are no longer isolated clusters of finite size.

\begin{figure}[!ht]
$$
\epsfig{file=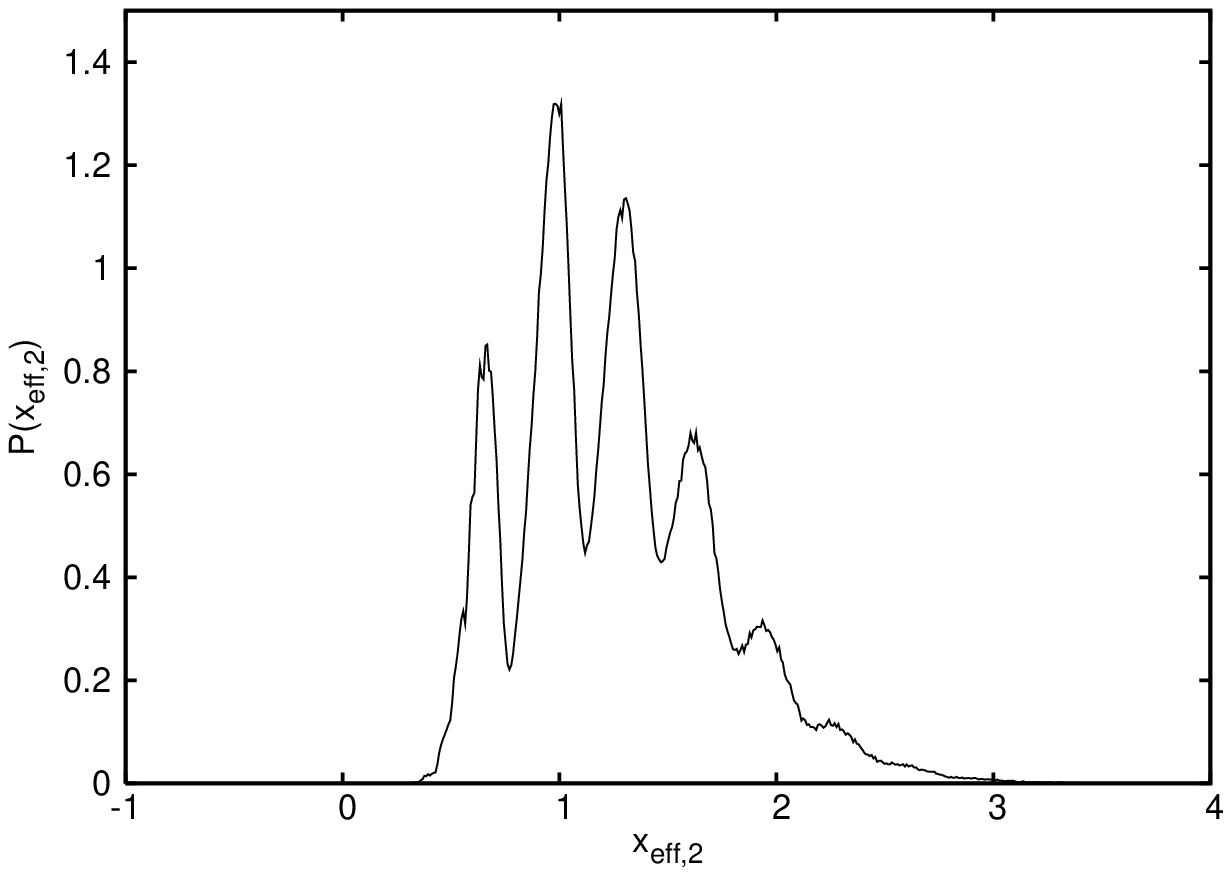,width=0.5\textwidth}
$$
\caption{Distribution $P(x_{{\rm eff},2})$ for a harmonically coupled system with homogeneous
coupling strengths $k_{ij}\ev 1$ on a non-Poissonian random graph of average connectivity
$C=4$ as described in the text; the inverse temperature is $\beta=1$.} \label{Fig:P_of_xeff2_Lg2}
\end{figure}

For the non-Poissonian random graph, a comparison with the distribution of the 
expansion coefficient $x_{{\rm eff},2}$ of the effective single site potentials in 
Fig \ref{Fig:P_of_xeff2_Lg2} reveals that these ensembles --- while qualitatively 
similar --- are indeed {\em not\/} identical.

\begin{figure}[!h]
$$
\epsfig{file=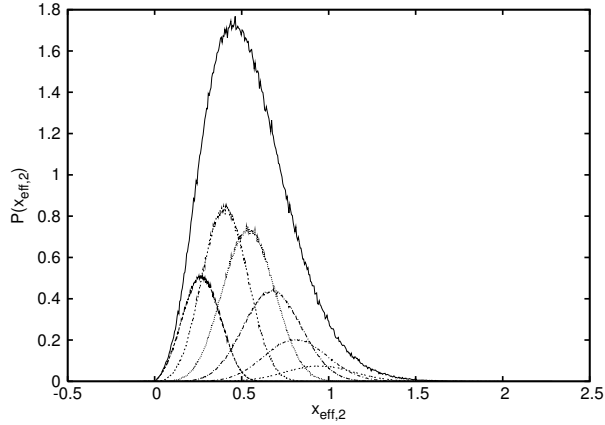,width=0.5\textwidth}
$$
\caption{Distribution $P(x_{{\rm eff},2})$ shown together with separate contributions to it 
coming from sites with connectivities $L=2,3,\dots,7$, for a harmonically coupled system 
with couplings $k_{ij} \sim \cU[0,1]$ on a non-Poissonian random graph of average 
connectivity $C=4$ as described in the text; the inverse temperature is $\beta=1$.} 
\label{Fig:P_of_xeff2_Lg2_Var_K_L2-7}
\end{figure}

As a last result for the harmonic system, we look at the distribution  the 
expansion coefficient $x_{{\rm eff},2}$ for a system in which we have non-uniform coupling
strengths in addition to locally varying connectivities. In the present case we took $k_{ij}$
{\em uniformly\/} distributed in [0,1], and choose a non-Poissonian graph structure of with
$C=4$ of the same type as before. Due to the variability in the $k_{ij}$, the peak structure
in the $x_{{\rm eff},2}$-distribution is smeared, and it is no longer possible to associate
a value of the localization length with a local connectivity, simply by looking at the 
distribution. However, the method of determining the distribution of effective single site 
potentials easily allows to monitor the contributions coming from sites with different 
connectivities, as illustrated in Fig. \ref{Fig:P_of_xeff2_Lg2_Var_K_L2-7}.

\section{Glassy Sytems with Anharmonicities}			\label{sec:anharm}

The next model that we investigate is a random graph of  coupled
(scalar) degrees of freedom with a distribution $P(k_{ij})$ of harmonic
coupling strengths and fixed anharmonic coupling strength. Furthermore, we
assume that there are non-random anharmonic single-site potentials.
The interaction energy between connected vertices $(i,j)$ of the
graph is taken to be of the form
\beq
\phi_{ij}(u_i-u_j)={k_{ij}\ov2}(u_i-u_j)^2+\la_4(u_i-u_j)^4
\eeq
and we introduce a vertex-independent single-site potential 
\beq
V(u_i)=g_4u_i^4
\eeq

If the support of $P(k_{ij})$ contains negative couplings, at least one of
the quartic couplings $\la_4$ or $g_4$ must be positive for the system to be 
stable. 

A fully connected variant of this model with $g_4 >0$ and $\la_4  =0$ was studied
in \cite{KuUr00}, the corresponding model with $g_4 =0$ and $\la_4  >0$  in 
\cite{Kuehn03}. For the purposes of describing glassy low-temperature anomalies
either form of stabilization by a quartic anharmonicity appeared to be acceptable.
Differences between low-temperature properties of the two models were mainly in the 
details, although from a fundamental point of view the system with $g_4=0$ would 
of course be preferable, since $g_4 \ne 0$ breaks translational invariance.

For the purposes of the present investigation we shall nevertheless stick to
the local stabilizing potentials, i.e. choose $\la_4  =0$ and $g_4 \ne 0$, because
this form of stabilization is numerically much easier to handle than its 
translationally invariant counterpart. We will turn to anharmonic systems with 
full translational invariance in a separate publication.

From the solutions for the distributions $\pi[\psi]$ and $\hpi[\hps]$ of single-replica
potentials one obtains a distribution of effective-single site potentials via (\ref{PVeff}).

Using an expansion of effective single-site potentials $V_{\rm eff}(u)$ in terms of
the modified Hermite polynomials used in Sec \ref{sec:harm} to describe harmonic 
systems, 
\beq
V_{\rm eff}(u) = \sum_\mu x_\mu \varphi_\mu(u)\ ,
\eeq
the distribution of effective-single site potentials translates into a distribution
of expansion coefficients $\{x_\mu\}$ (to simplify notation, we drop the subscript 
`eff' on the expansion coefficients).

For systems with (quartic) anharmonicities, the solution of the fixed point equations
(\ref{SP_hpi}), (\ref{SP_pi}) involves distributions of single-replica potentials 
$\psi$ and $\hps$ with support on potentials having anharmonicities of {\em all\/} 
orders. Any finitely truncated orthogonal function representation of these 
distributions would therefore only generate an approximate solution to the full problem
(on top of the approximation implied in the RS assumption). It is therefore advisable 
to use the full functional approach (not least in order to check the quality of finite 
dimensional approximations).  Below we therefore show distributions of some low order
expansion coefficients as obtained in the functional approach and compare them with 
corresponding distributions obtained within a low-order truncation scheme.

We fix the energy scale of the system by setting $g_4=1$, and introduce frustration
in the system by taking Gaussian distributed harmonic couplings, $k_{ij} \sim 
\cN(0,\sigma_k)$. A system of this type then undergoes a transition from a symmetric
high temperature phase, in which all effective single-site potentials respect the
microscopic inversion symmetry of the original interaction energy $V_{\rm eff}(u)=
V_{\rm eff}(-u)$, to a glassy low-temperature phase, in which this symmetry is 
spontaneously broken, i.e. which is described by an ensemble of single-site potentials
for which {\em typically\/} $V_{\rm eff}(u) \ne V_{\rm eff}(-u)$. This transition
occurs at a temperature $T_c$ which depends on the strength $\sigma_k$ of the 
disorder (and of course on the connectivity distribution). Indeed, for a combination
of quadratic and quartic potentials as chosen in the present setup, dimensional analysis
reveals that the critical condition should have a parameter dependence of the form 
\beq
T_c(\sigma_k, g_4) =\frac{\sigma_k^2}{g_4} T_c(1,1)\ .
\label{Tc_of_sig}
\eeq
as already found for the fully connected system \cite{Kuehn03}. Here the value of 
$T_c(1,1)$ still depends, of course, on the connectivity distribution. E.g., for 
the system with the non-Poissonian connectivity distribution with average degree 
$C=4$ looked at earlier, we determined $T_c(1,1)=0.932 \pm 0.005  \Leftrightarrow
\beta_c(1,1) \simeq 1.073$, and we verified the scaling (\ref{Tc_of_sig}) with high 
precision within the full functional treatment of the self-consistency equations. 
It is important to note, however, that approximate treatments, such as the low-order 
orthogonal function approximation procedure, need not respect this scaling of  
$T_c(\sigma_k, g_4)$, though we found them to be not far off.

\begin{figure}[!h]
$$
\epsfig{file=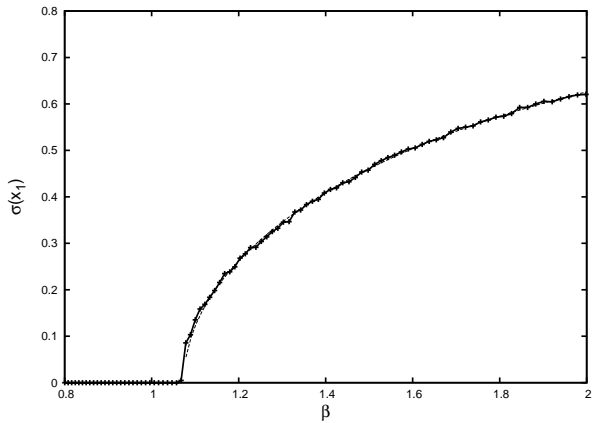,width=0.5\textwidth}
$$
\caption{Width $\sigma(x_1)$ of the distribution of the first expansion coefficient $x_1$
in in the ensemble of effective single-site potentials. The width goes to zero with a 
square-root singularity at $\beta_c \simeq 1.073$. A fit displaying a square-root singularity
with analytic corrections describes the data very well over a wide range of inverse 
temperatures. For this system $\sigma_k=1$.} 
\label{Fig:sig_x1_of_beta_sigk1}
\end{figure}

The transition to glassy order at low temperatures is continuous in the sense that 
widths of the distributions of the  expansion coefficients $x_1$, $x_3$, $x_5$, \dots 
coupling to anti-symmetric functions go to zero continuously as the transition 
temperature is approached from below. Fig \ref{Fig:sig_x1_of_beta_sigk1} illustrates 
this for a system with a non-Poissonian connectivity distribution of the type used 
in Sec \ref{sec:harm}  with $C=4$, and $\sigma_k=1$. If an orthogonal function
representation of the single-replica functions in terms of Hermite polynomials, 
truncated at order four is used instead of the full functional approach, we find 
the order of the transition to be unchanged, though the transition point is shifted 
to a sligthly higher temperature, $T_c(1,1) \simeq 1.043 \Leftrightarrow  \beta_c(1,1)  
\simeq 0.959 $ (differing from the exact value by approximately 3\%).

We note that the population-dynamics algorithm exhibits critical slowing down as
the transition is approached. In Fig \ref{Fig:sig_x1_of_beta_sigk1}. It is thererfore
crucial to use sufficiently long runs and small temperature steps close to the transition 
to ensure that the algorithm has properly converged, before taking measurements.

\begin{figure}[p]
\vspace{-3.5cm}
\begin{center}
\epsfig{file=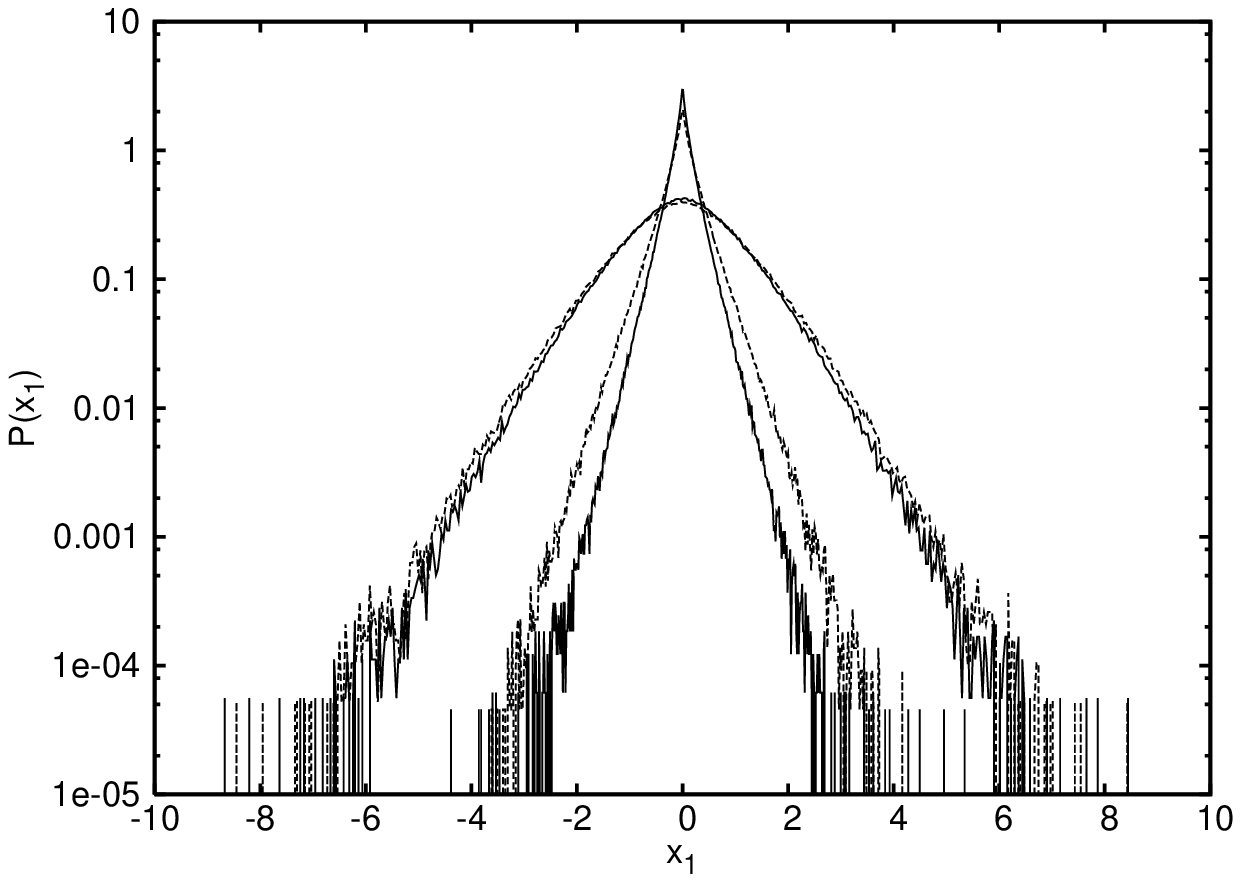,width=0.45\textwidth}\hfill
\epsfig{file=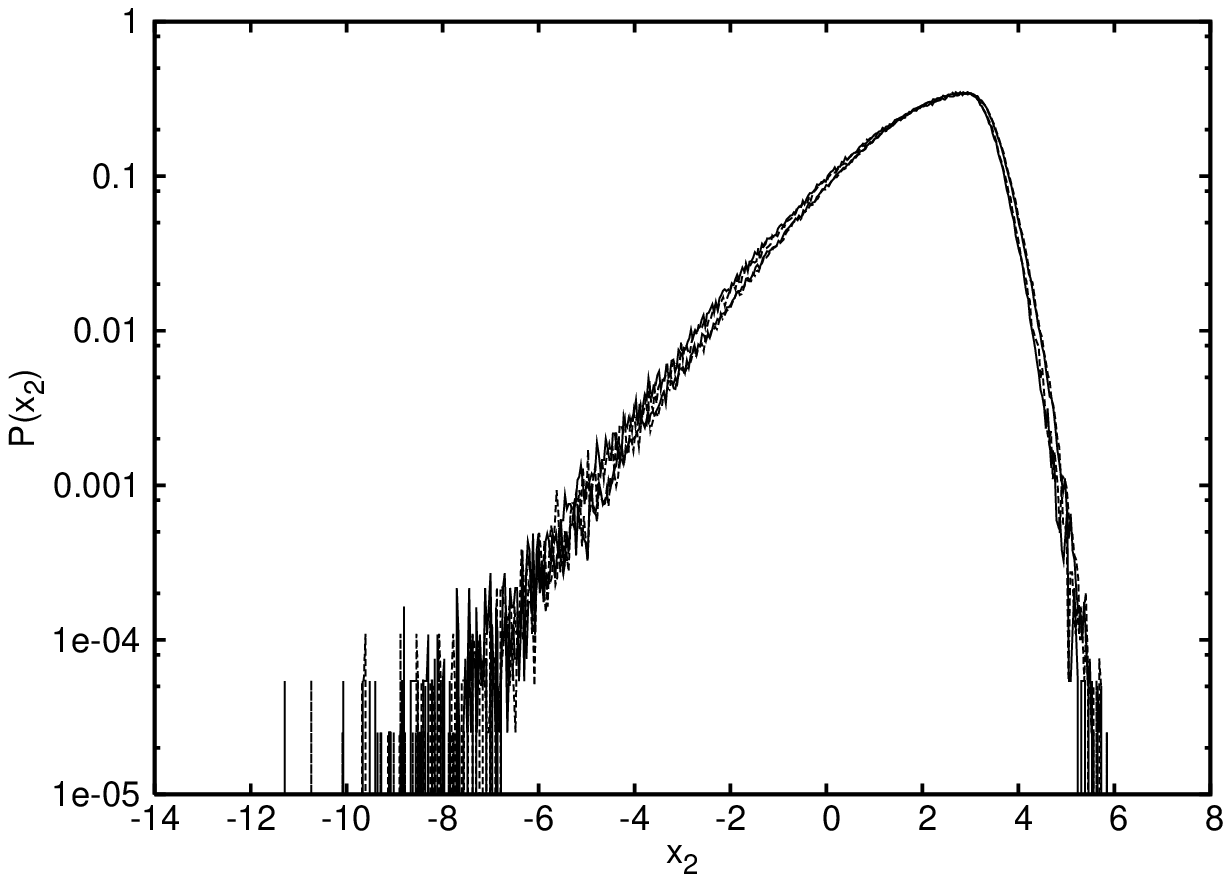,width=0.45\textwidth}
\epsfig{file=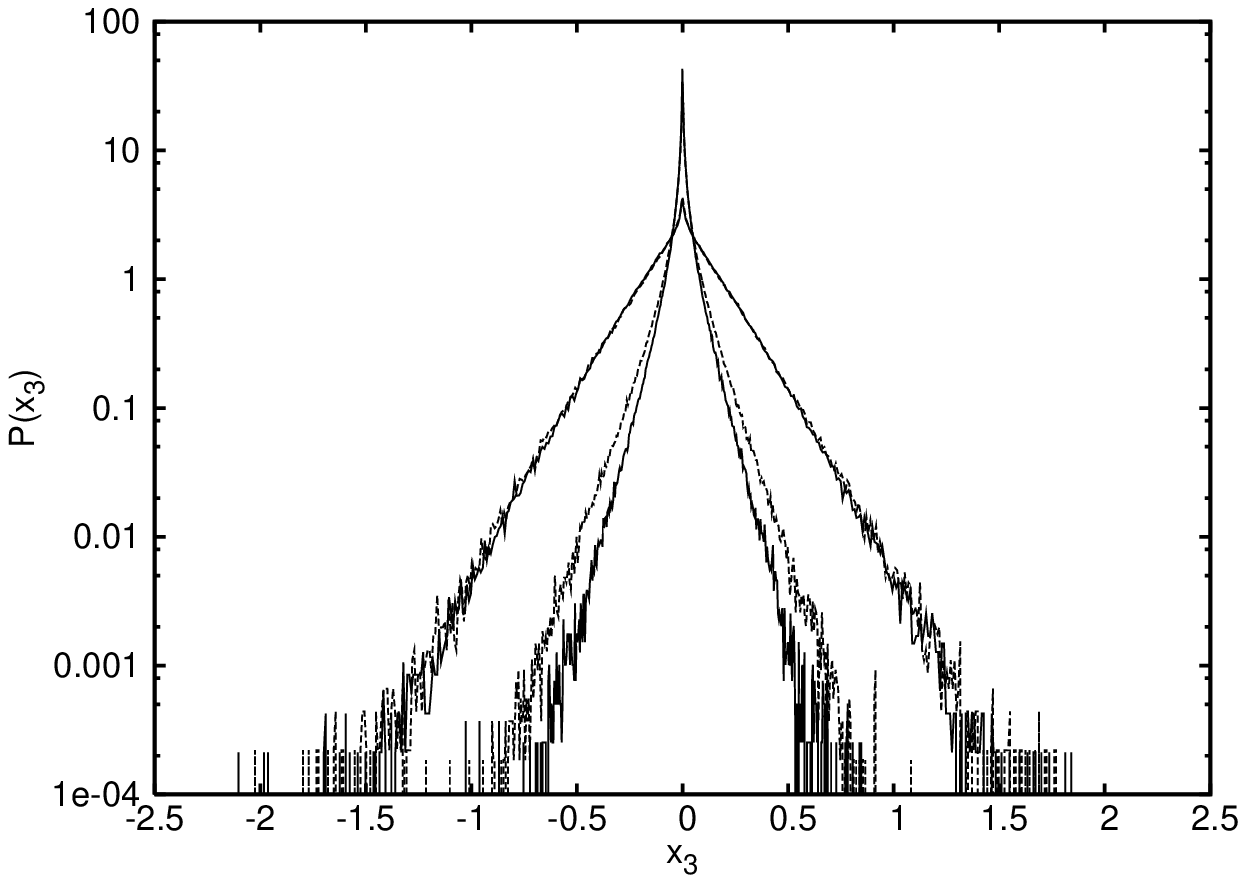,width=0.45\textwidth}\hfill
\epsfig{file=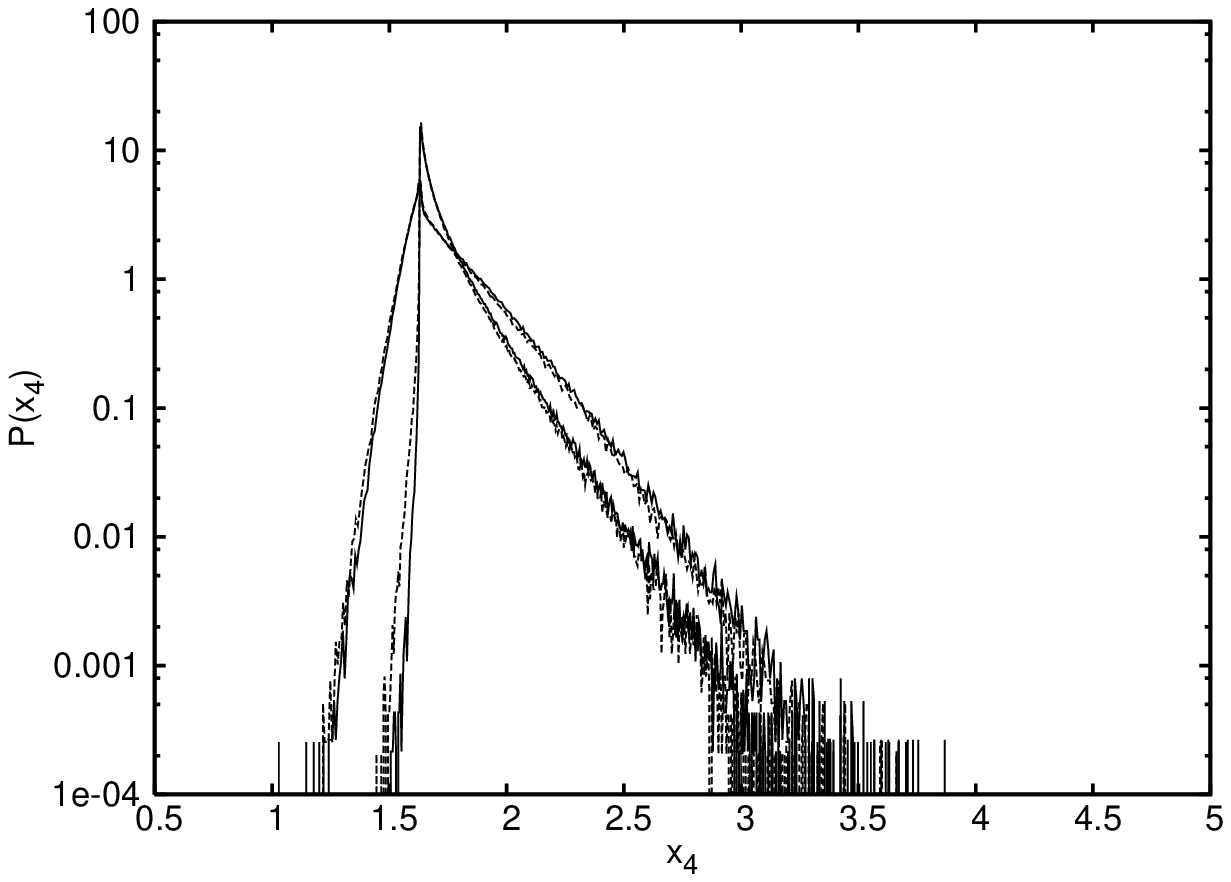,width=0.45\textwidth}
\epsfig{file=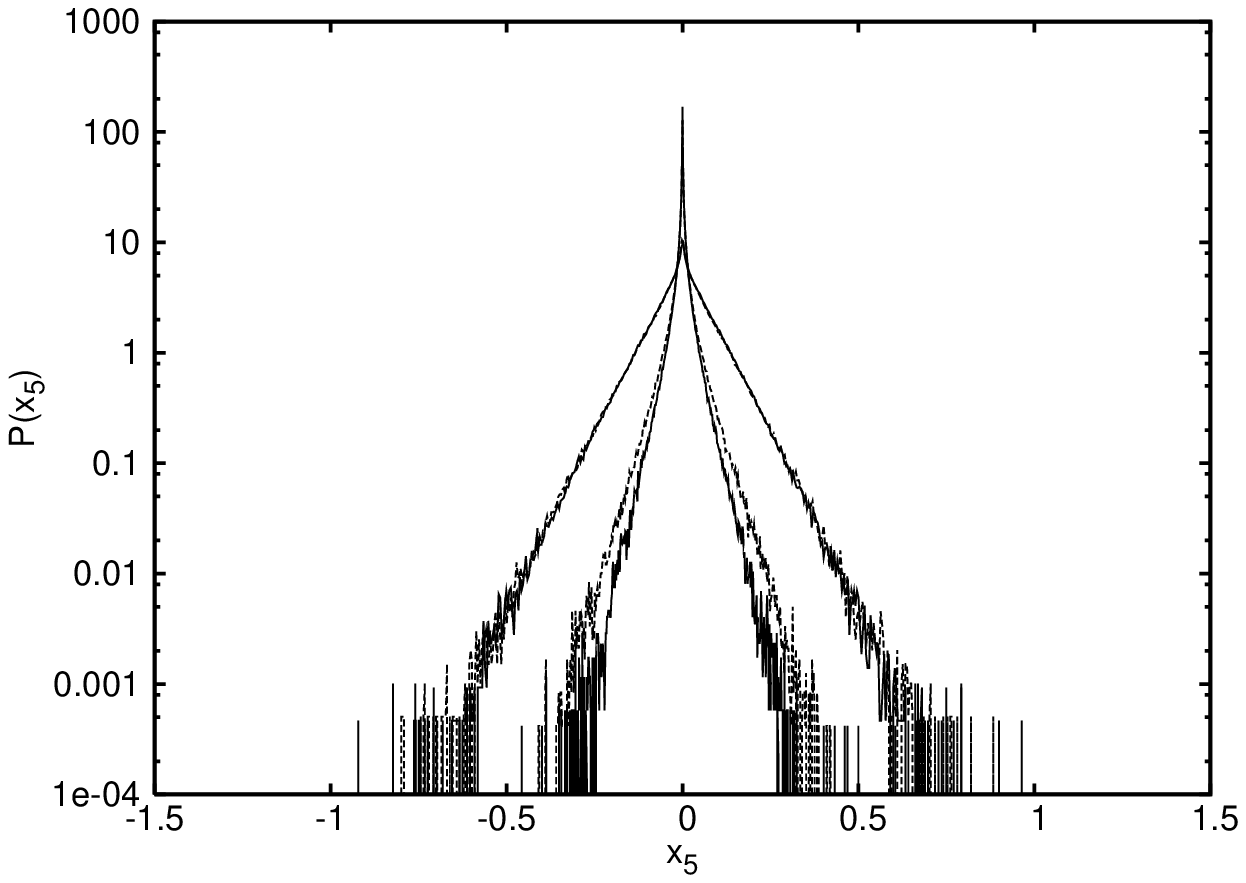,width=0.45\textwidth}\hfill
\epsfig{file=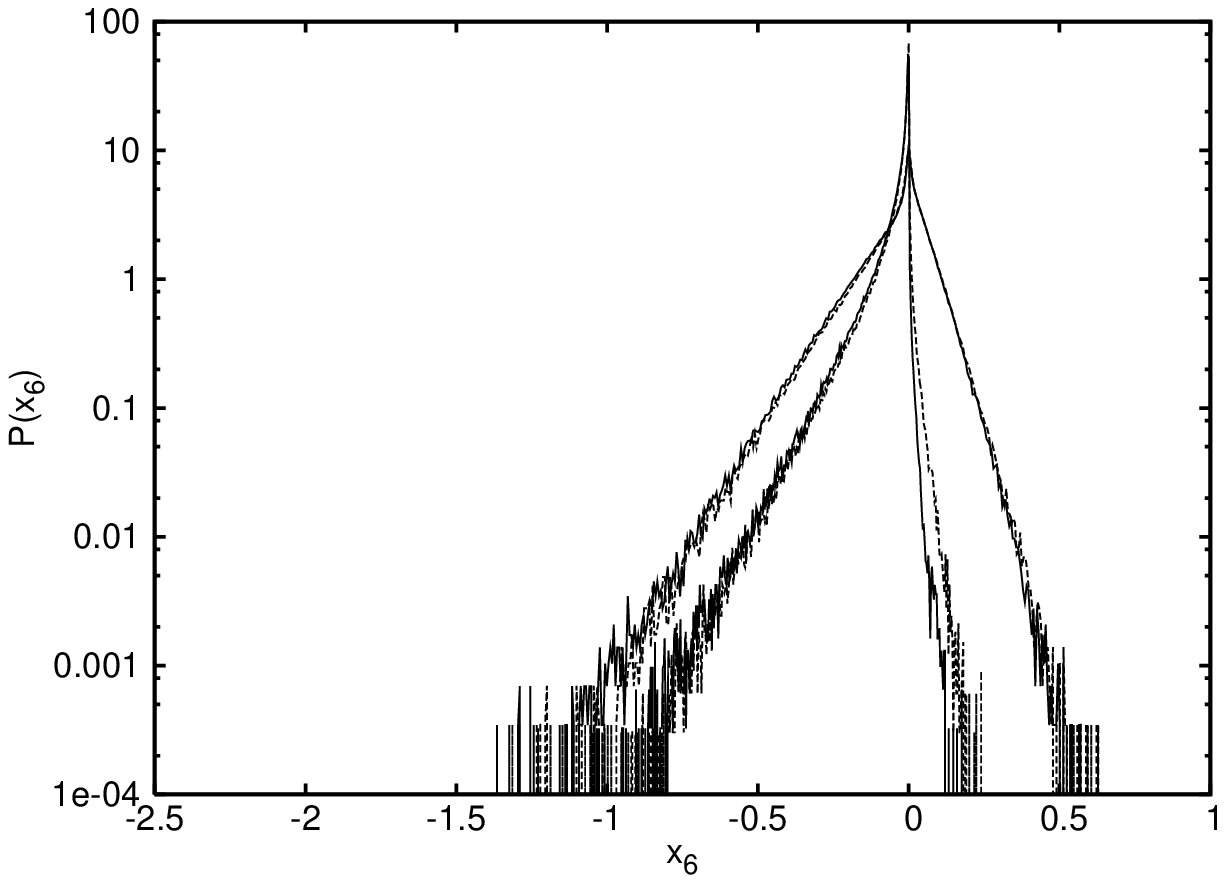,width=0.45\textwidth}
\epsfig{file=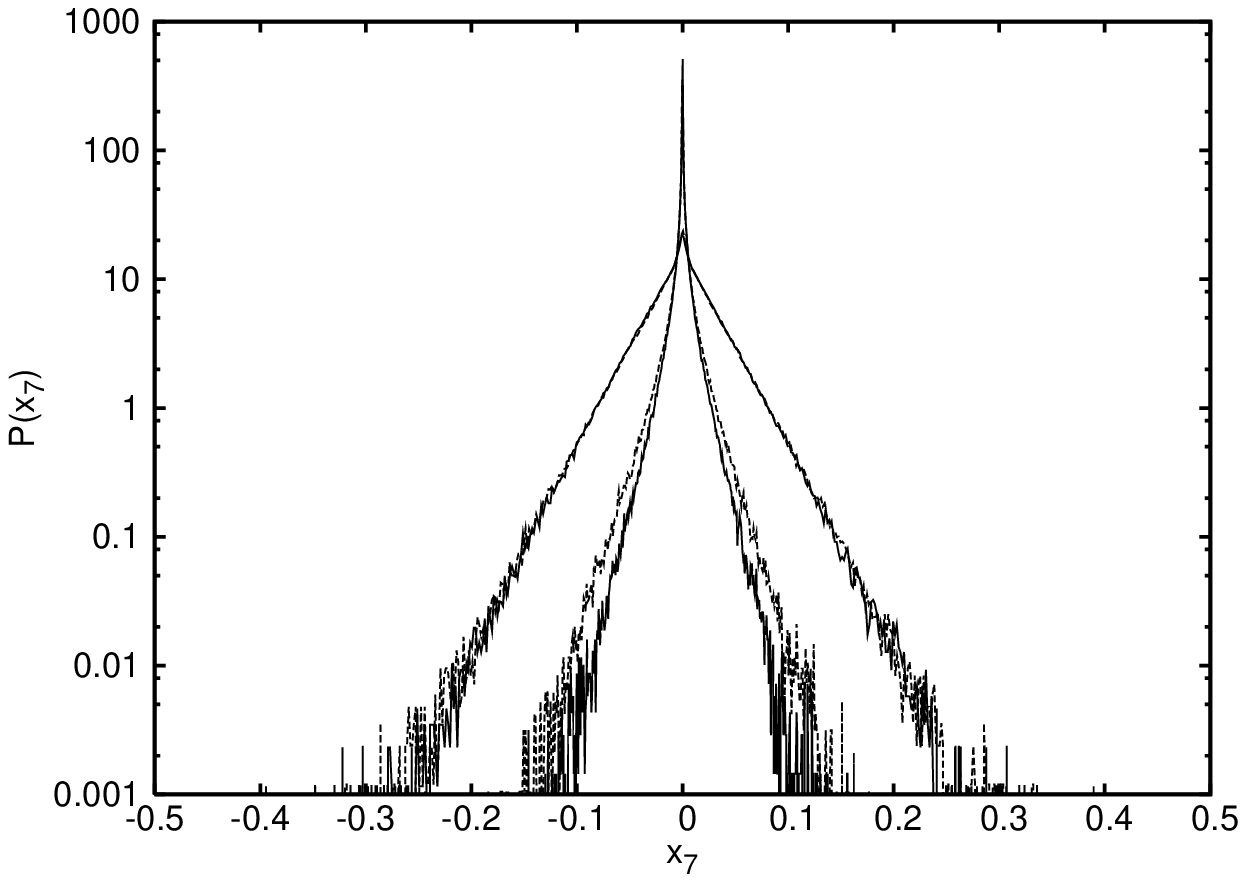,width=0.45\textwidth}\hfill
\epsfig{file=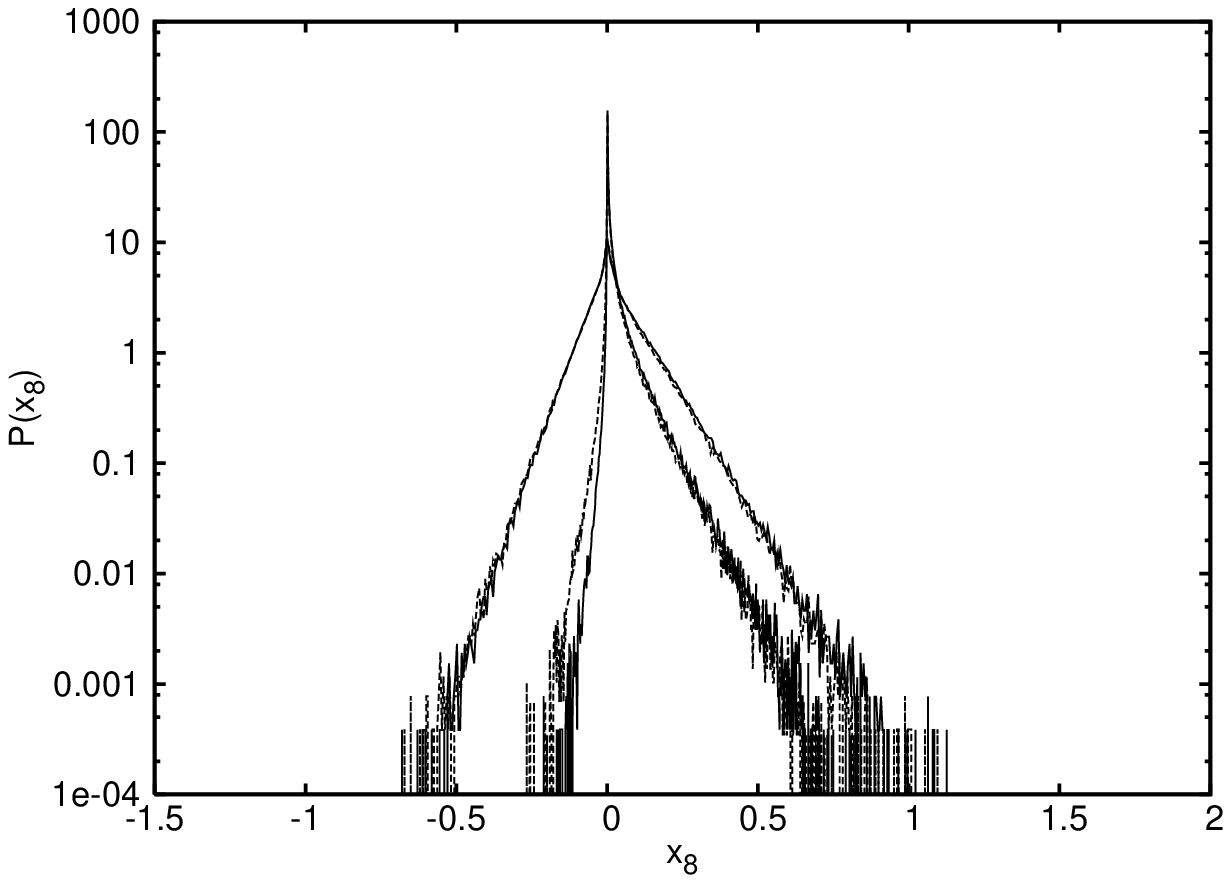,width=0.45\textwidth}
\end{center}
\caption{Distributions of low-order expansion coefficient $x_\mu$, $\mu=1,\dots,8$ 
in in the ensemble of effective single-site potentials for $\sigma_k=1.1$ at $T=1$
(narrower set of curves) and $T=0$ (wider set of curves). Solutions obtained within 
the full functional approach (full lines) are compared with those obtained
by truncating expansions of single-replica potentials in terms of Hermite polynomials 
at order 4. (dashed lines)}
\label{Pofxmu}
\end{figure}

Fig \ref{Pofxmu} shows distributions of expansion coefficients for a system with the 
non-Poissonian connectivity distribution used before, at $\sigma_k=1.1$, for $T=1$
and $T=0$, and compares distributions obtained from a full functional solution with 
those computed in an orthogonal function representation using Hermite polynomials that 
is truncated at order 4. Semilogarithmic plots are used to exhibit the behaviour in 
the tails of the distributions. At $T=1$, it turns out that the distributions of odd 
coefficients  $x_1$, $x_3$, \dots appear to be affected by the low order truncation 
in a slightly  stronger way than the distributions of even oder coefficients, and 
distributions of low order odd coefficients appear to be systematically slightly wider 
when computed via a truncated expansion approach; the main reason seems to be that the 
low order truncation has slightly higher critical temperature and with identical 
parameter settings, the system appears to be deeper in the glassy phase when described 
within the low order truncation scheme than when described in terms of full functional 
self-consistency. The distributions of the even coefficients, i.e. $x_2$, $x_4$ \dots 
computed in an approximation scheme truncated order 4 appear to be less strongly affected 
by the truncation scheme. 

\begin{figure}[p]
\vspace{-3.0cm}
\begin{center}
\epsfig{file=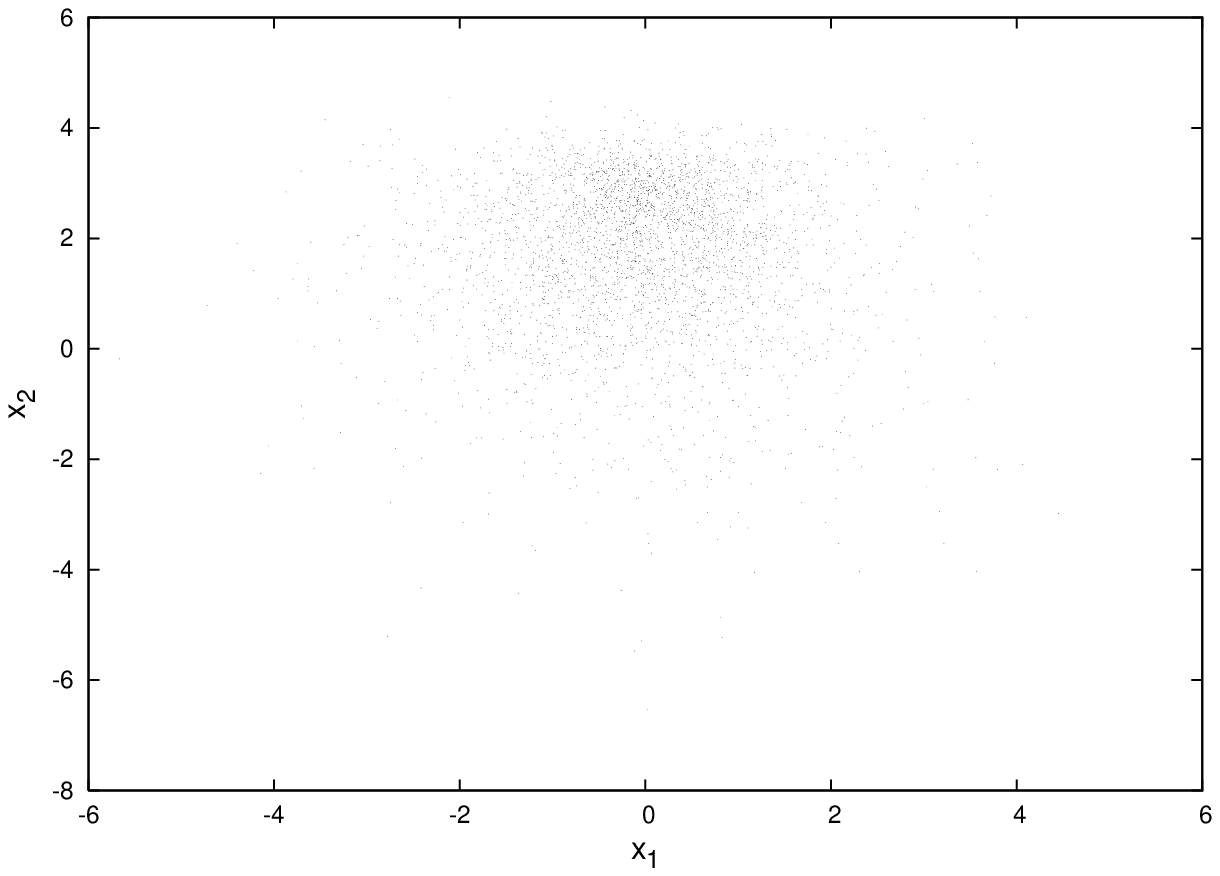,width=0.3\textwidth}
\epsfig{file=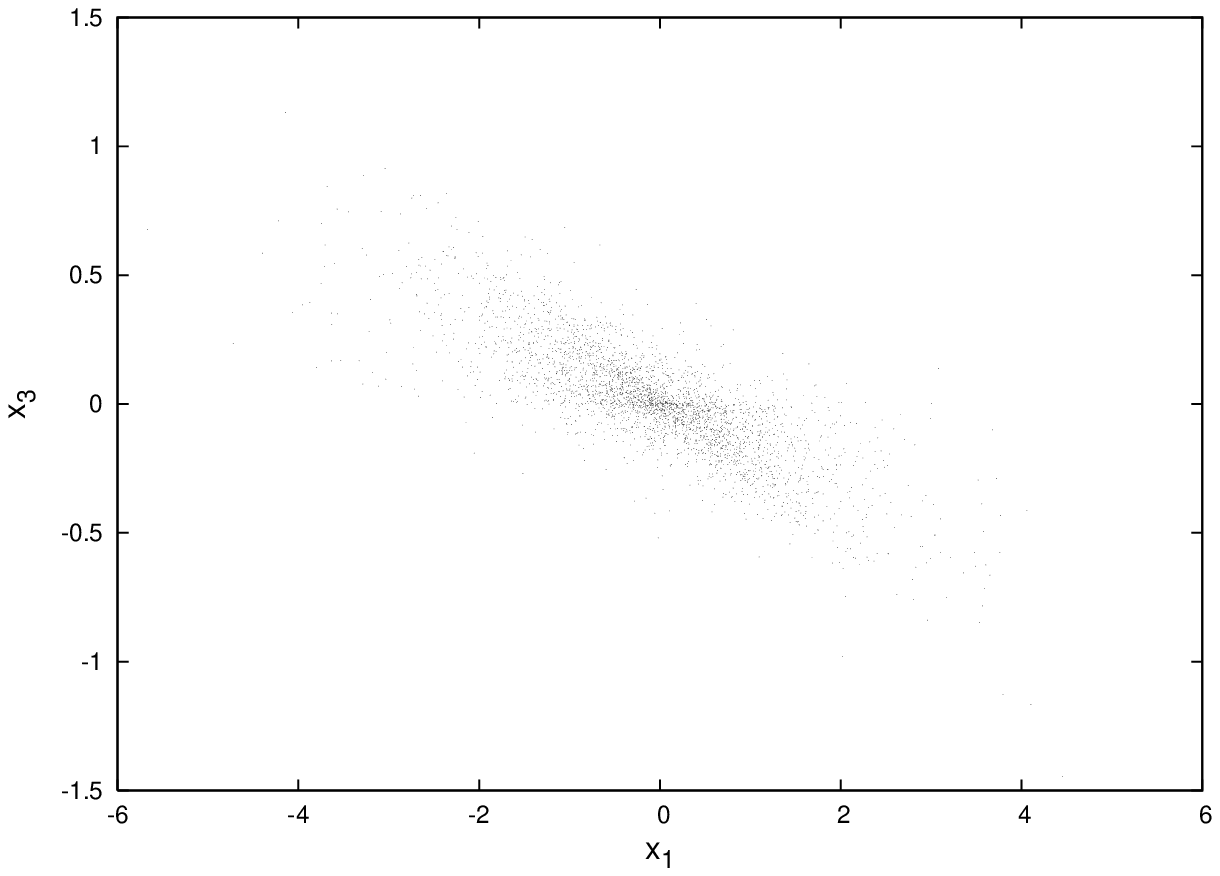,width=0.3\textwidth}
\epsfig{file=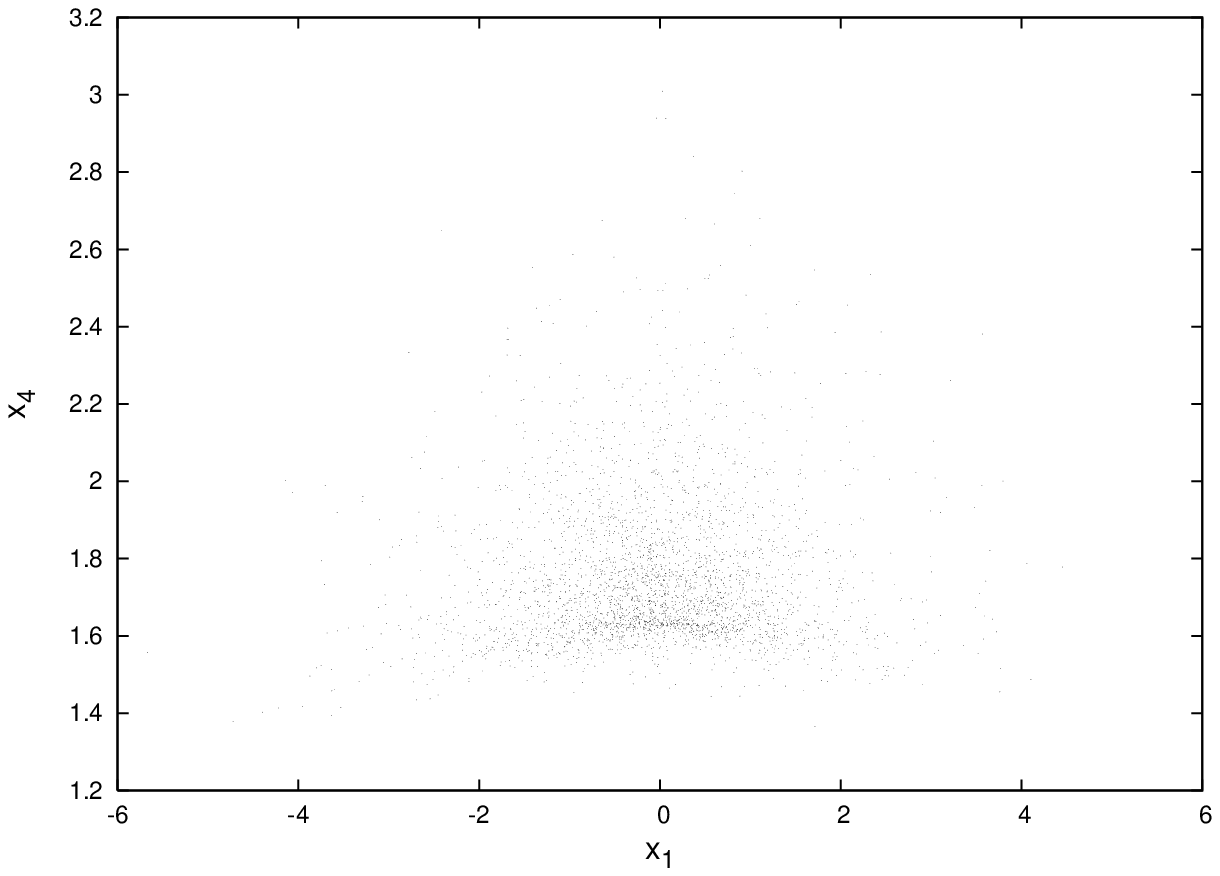,width=0.3\textwidth}
\\
\epsfig{file=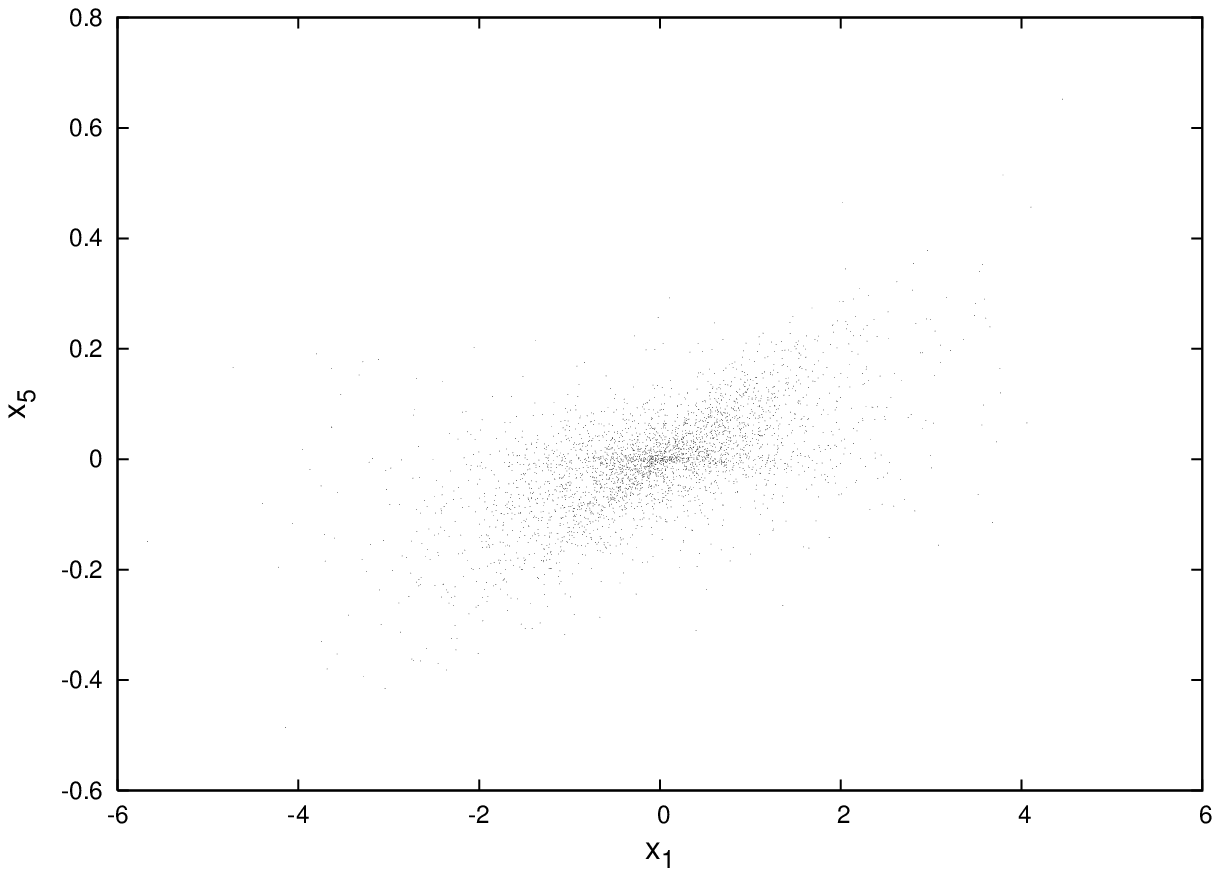,width=0.3\textwidth}
\epsfig{file=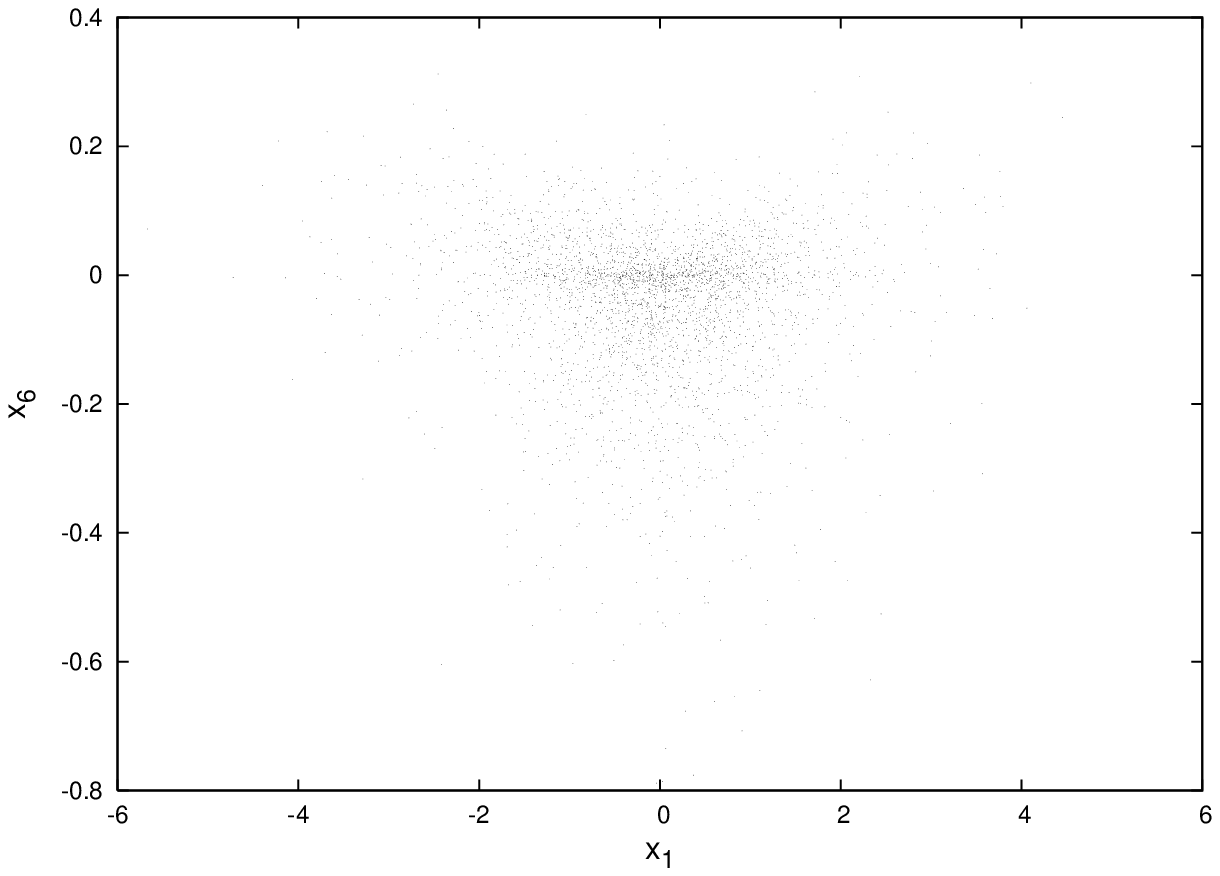,width=0.3\textwidth}
\epsfig{file=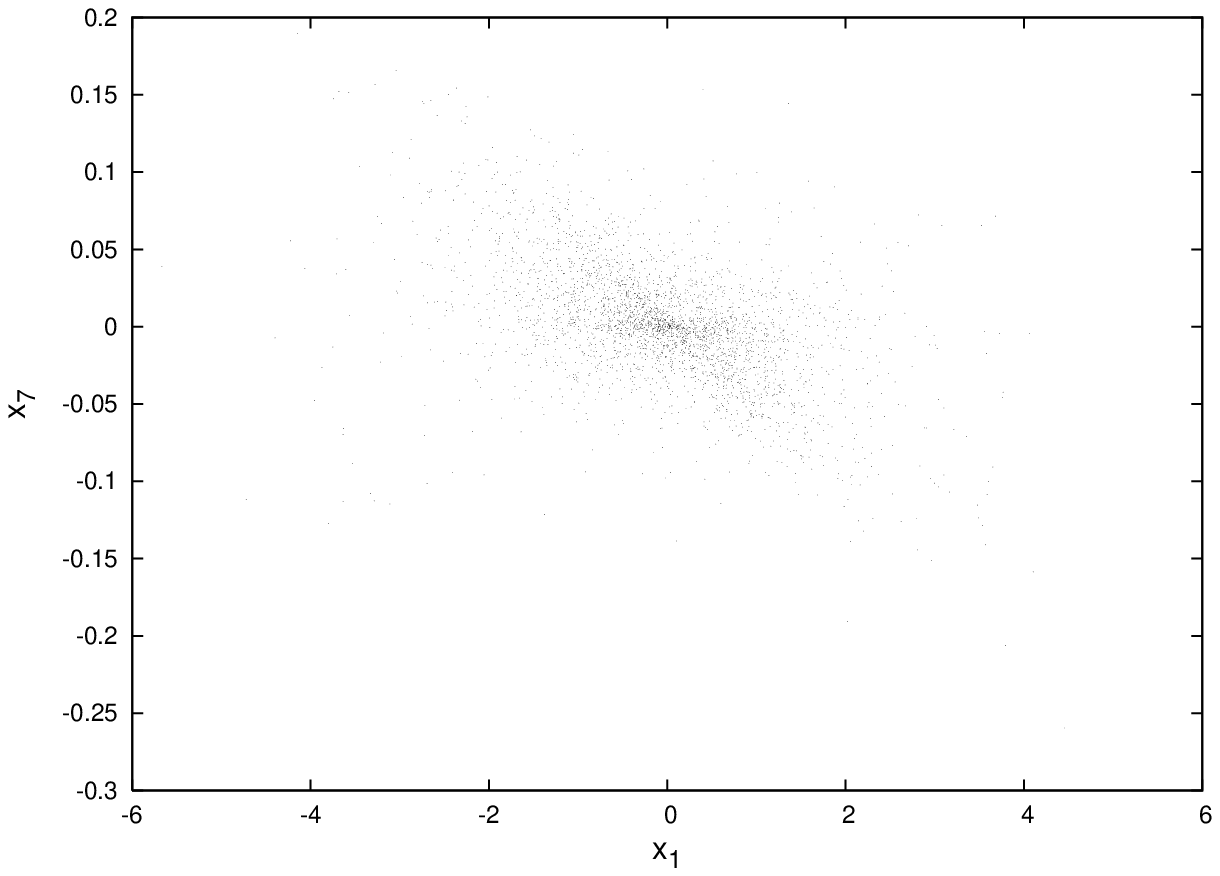,width=0.3\textwidth}
\\
\epsfig{file=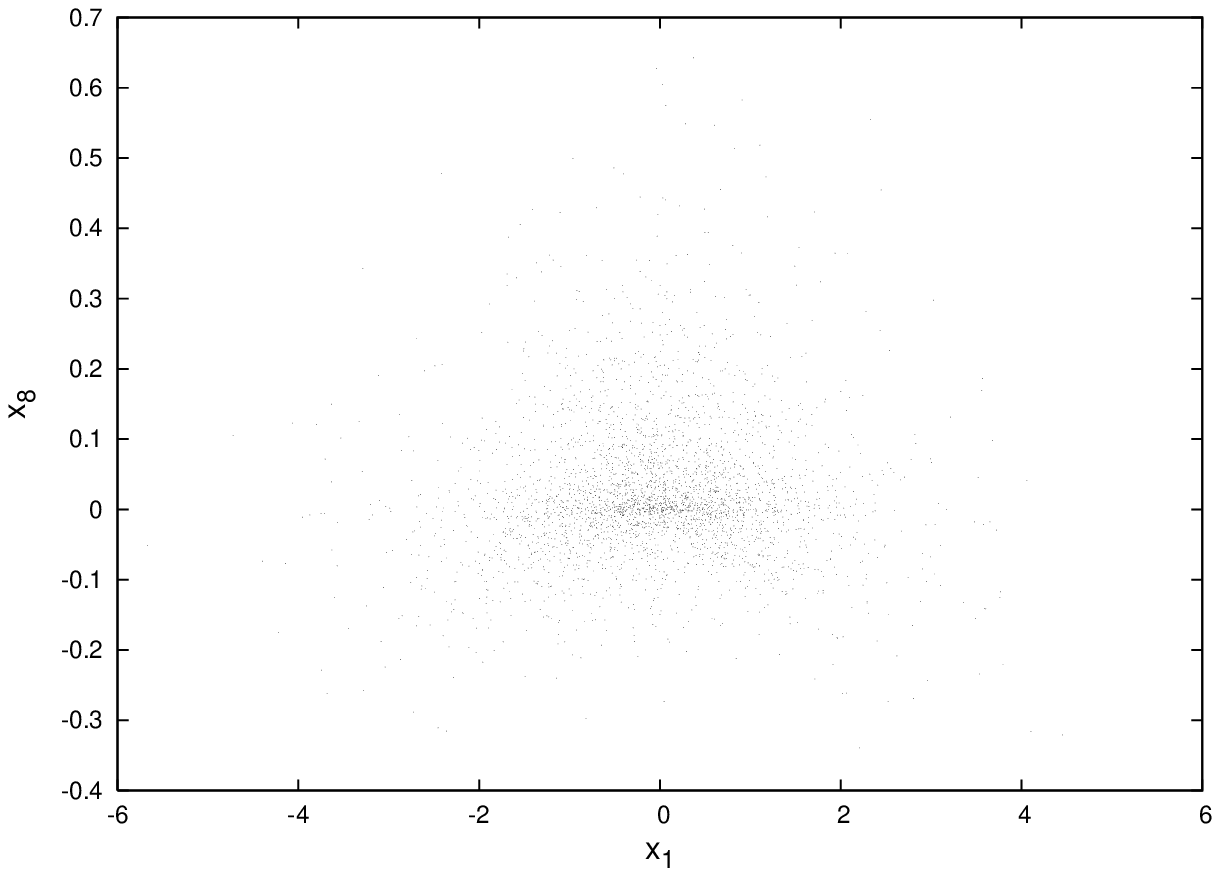,width=0.3\textwidth}
\\
\epsfig{file=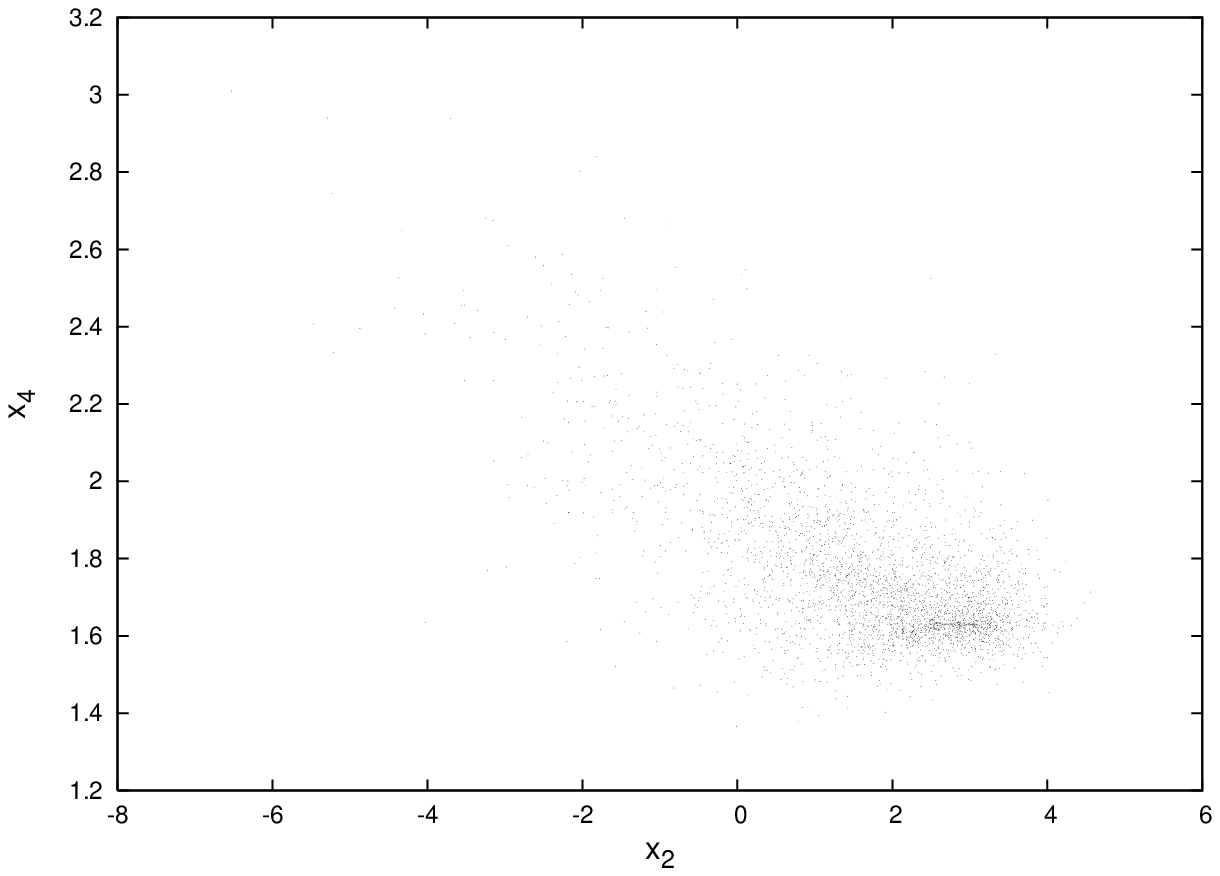,width=0.3\textwidth}
\epsfig{file=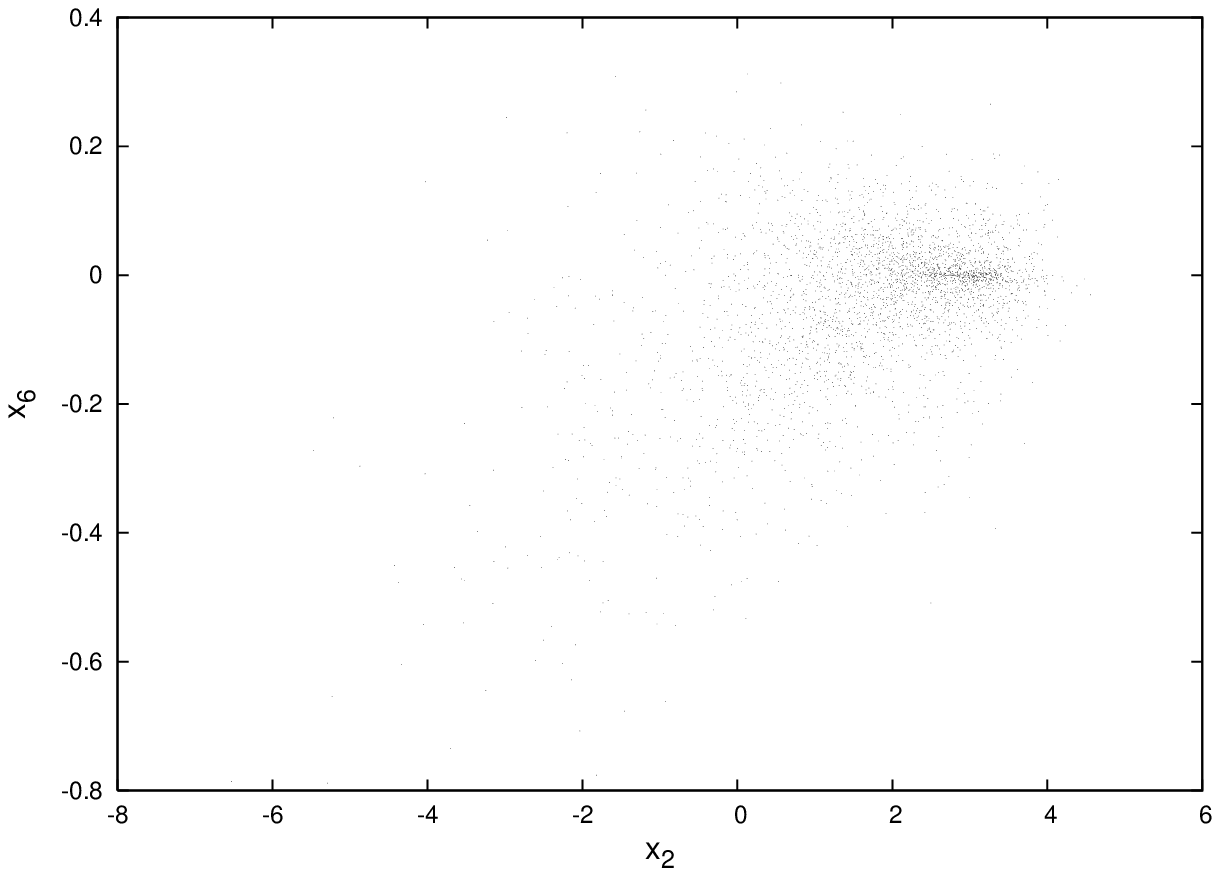,width=0.3\textwidth}
\epsfig{file=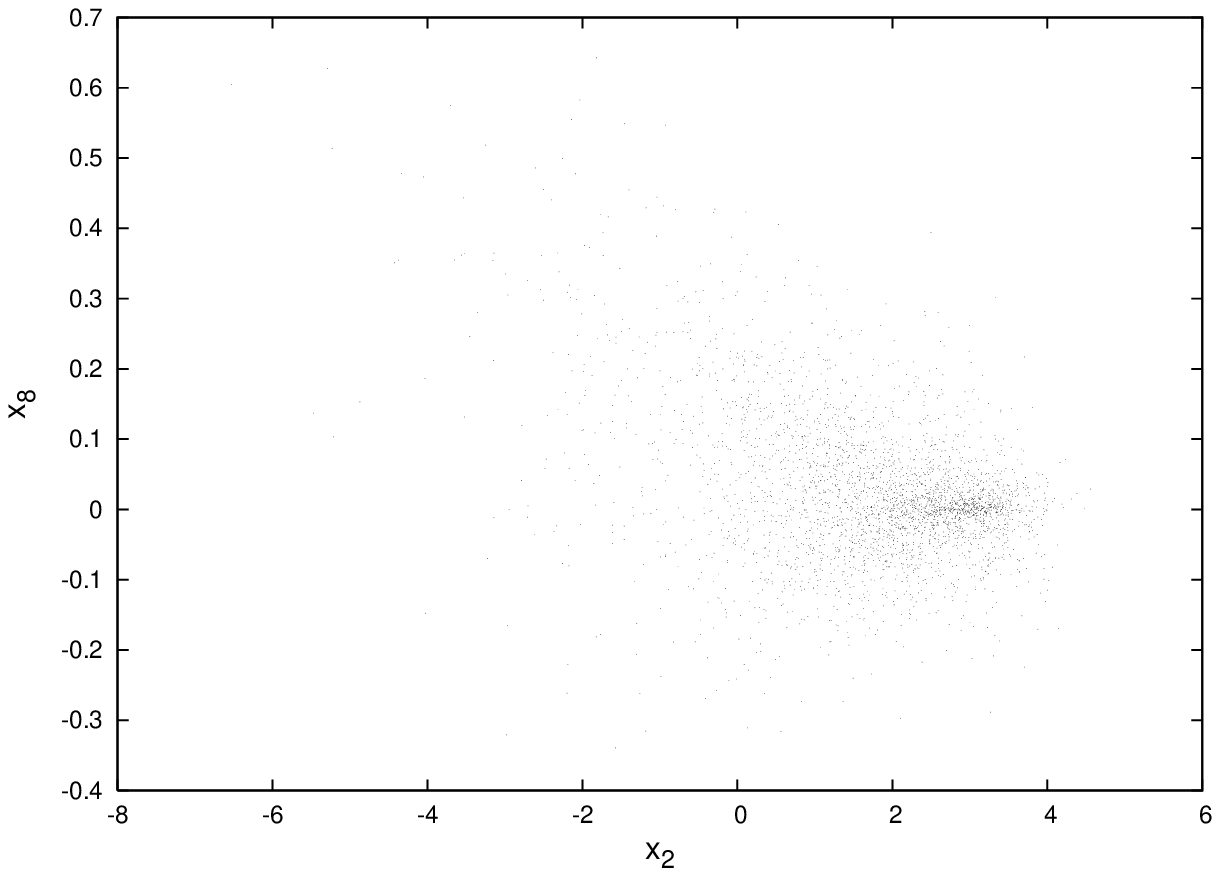,width=0.3\textwidth}
\\
\epsfig{file=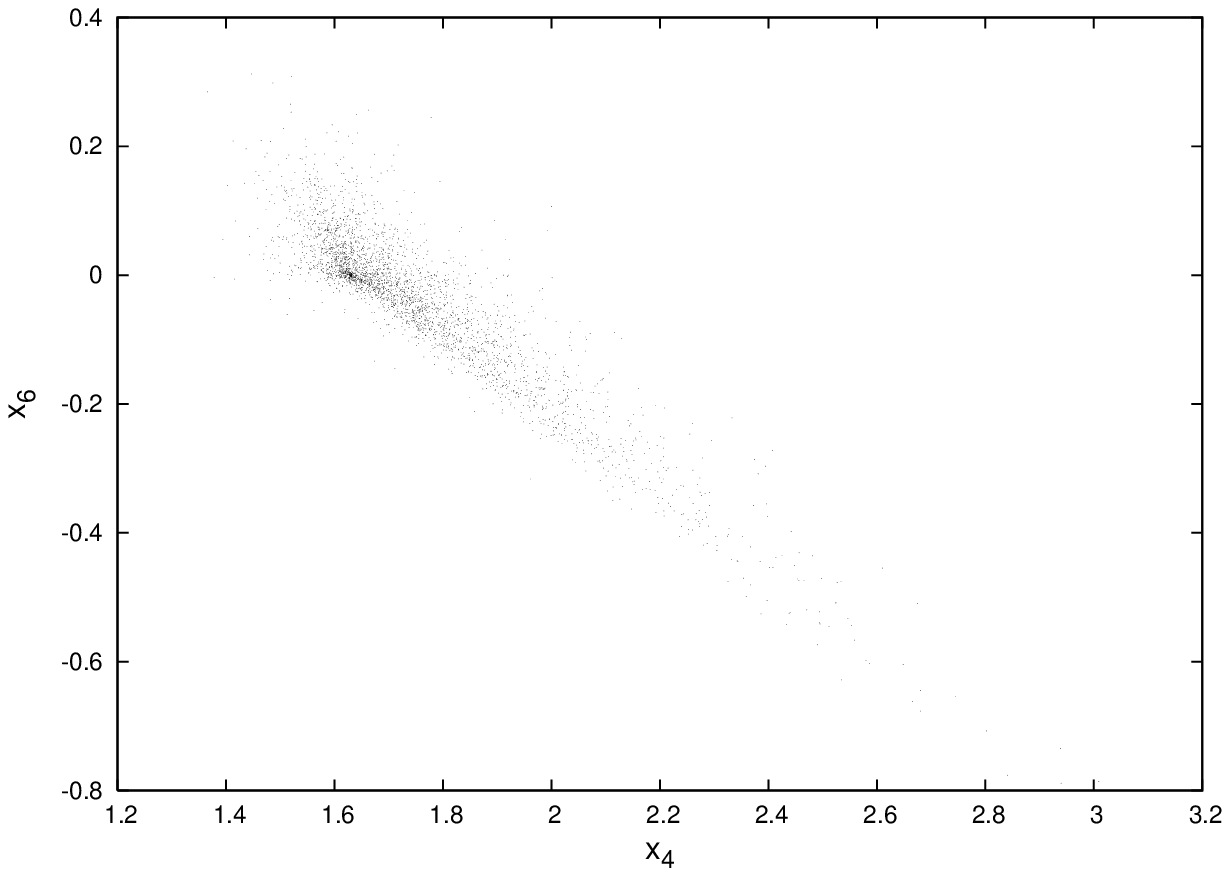,width=0.3\textwidth}
\epsfig{file=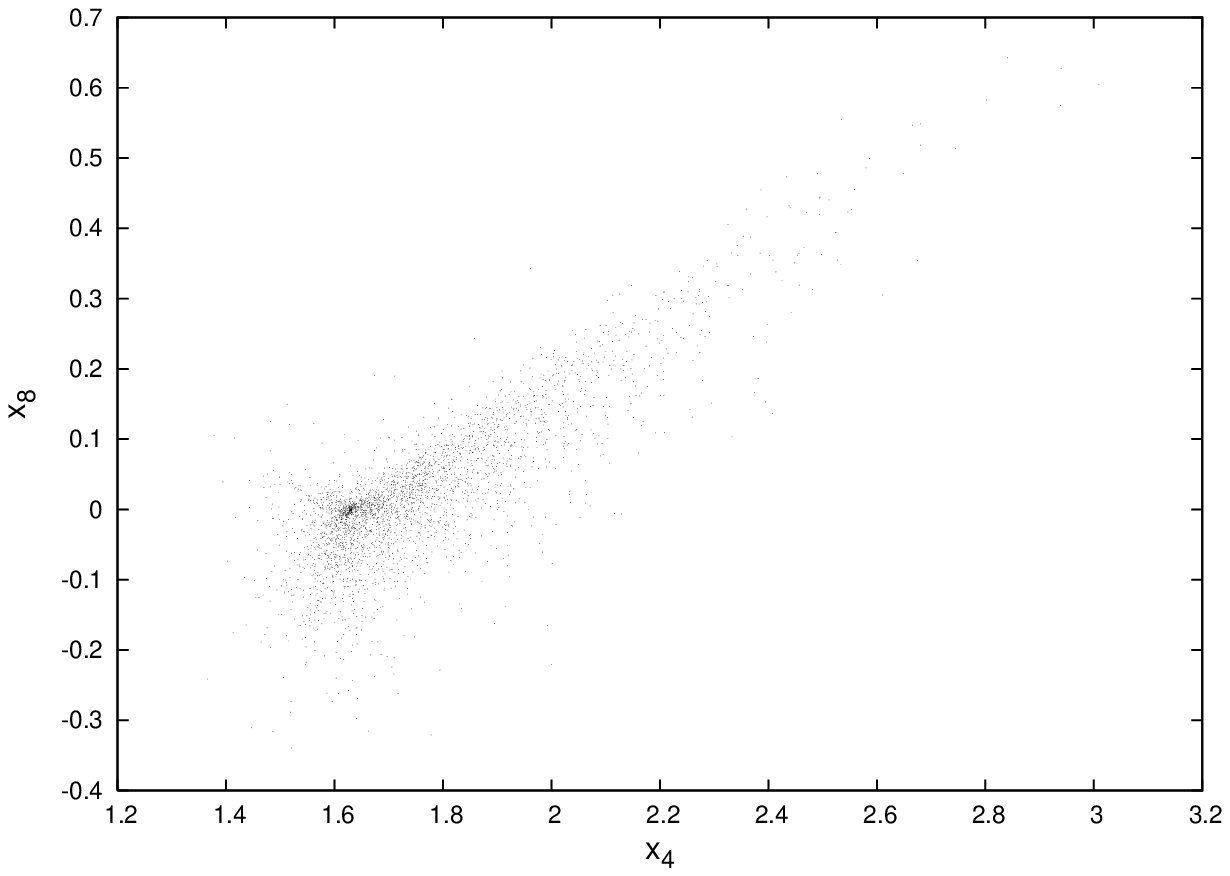,width=0.3\textwidth}
\\
\epsfig{file=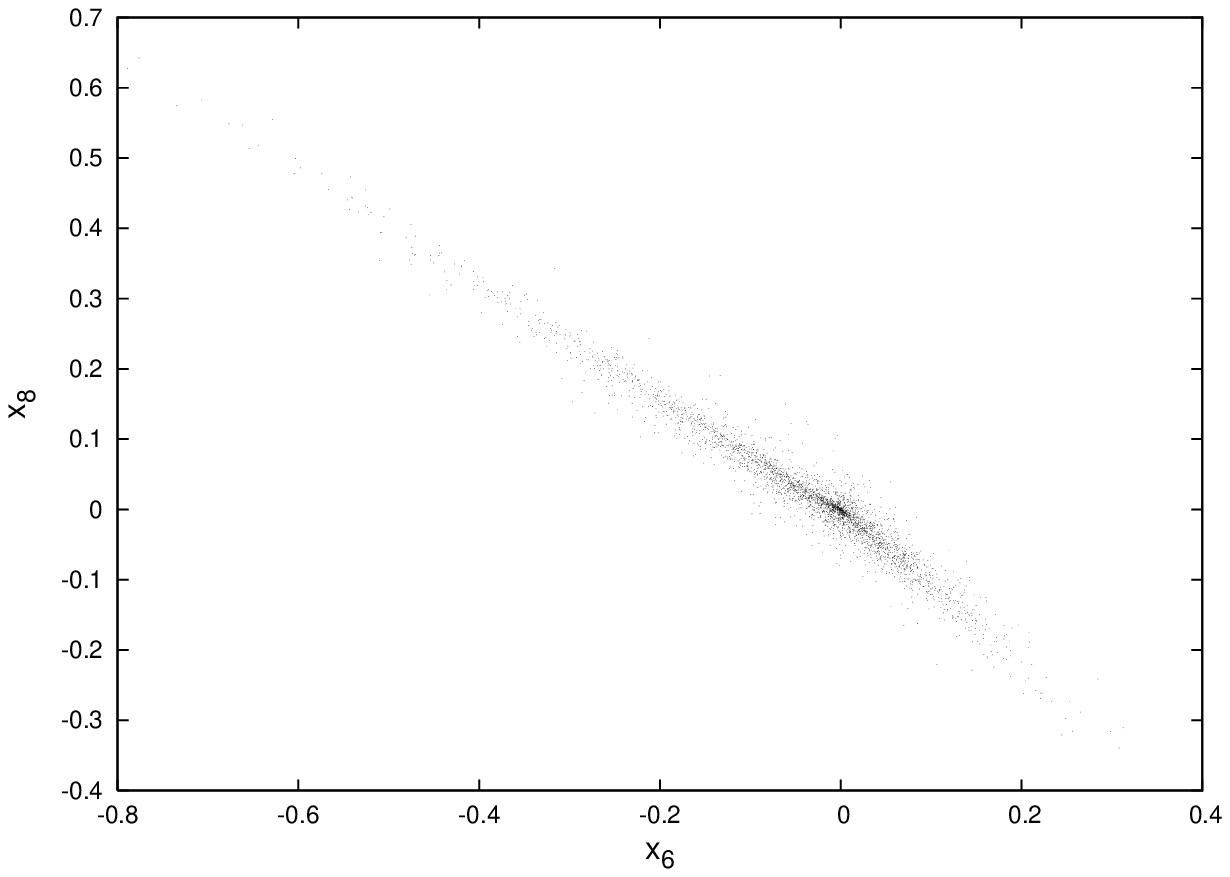,width=0.3\textwidth}
\end{center}
\caption{Scatterplots showing 2 dimensional projections of vectors of expansion 
coefficients for realizations of effective potentials. First group of panels $(x_1,x_2),
(x_1,x_3), \dots (x_1,x_8)$, second group $(x_2,x_4), (x_2,x_6), (x_2,x_8)$, third 
group  $(x_4,x_6)$ and $(x_4,x_8)$, and last panel $(x_6,x_8)$.
}
\label{scatterplots2D}
\end{figure}

It should be noted that there are strong correlations between the expansion coefficients
that can be exhibited, e.g. by computing their correlation coefficients or by inspecting
scatterplots of individual realizations. Fig. \ref{scatterplots2D} shows some examples.
For the system under investigation we found the $(x_{2k-1},x_{2\ell})$ scatterplots 
to be similar up to scaling and symmetry transformations; analogous similarities exist in
$(x_{2k},x_{2\ell})$-families at fixed $k$.

Differences between the full functional solutions and those obtained within a low-order 
expansion become rapidly less pronounced when moving away from the transition temperature.
This is particularly clear in the the zero-temperature limit where an orthogonal function 
representation truncated at order 4 describes the results remarkably well, as seen in the
wider set of curves also displayed in Fig \ref{Pofxmu}. This might have been anticipated, 
as anharmonicities typically play a less significant role at low temperatures.

Another feature that deserves mention is that the distribution $P(x_2)$ of the second
order coefficient his hardly temperature dependent at all (e.g., the zero-temperature limit
of this distribution is virtually indistinguishable from the one at $T=1$ (as seen in the
second panel of Fig \ref{Pofxmu}) whereas all other distributions do show a noticeable 
variation with temperature.

\begin{figure}[!h]
$$
\epsfig{file=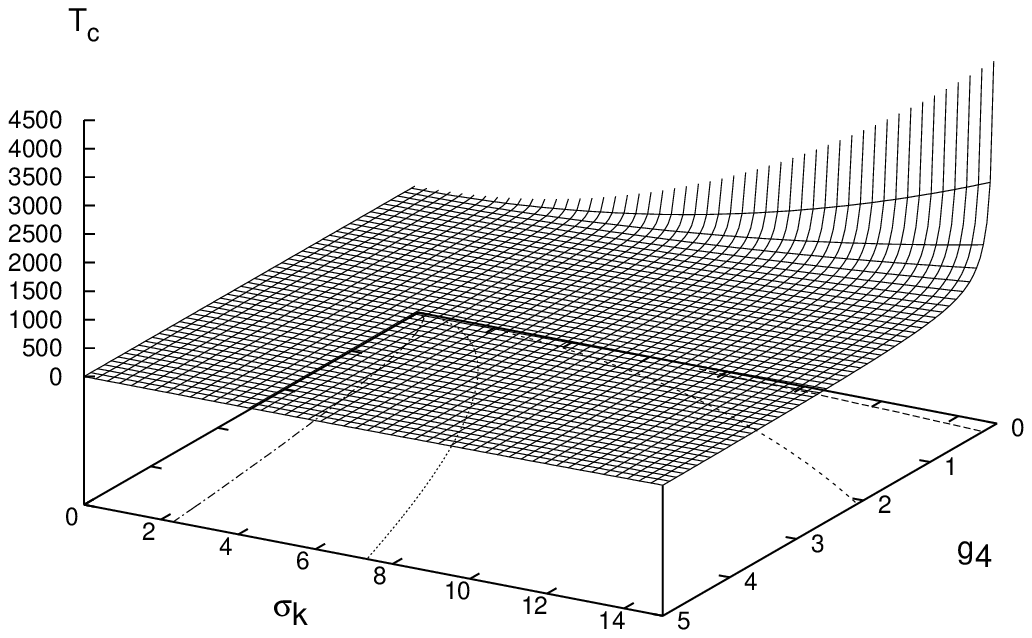,width=0.75\textwidth}
$$
\caption{Transition temperature of the anharmonic system as a function of the standard
deviation $\sigma_k$ of the random harmonic coupling, and the strength $g_4$ of the quartic
stabilizing potential. Iso-lines for $T_c=0.1$, $T_c=1$, $T_c=10$, $T_c=100$, and $T_c=1000$
are also shown in the $(\sigma_k,g_4)$-plane} 
\label{Fig:phasediagr}
\end{figure}

Fig \ref{Fig:phasediagr} gives the phase diagram of the system, showing the 
transition temperature as a function of the strength $\sigma_k$ of the disorder in 
the harmonic coupling constants.

\begin{figure}[ht]
\begin{center}
\epsfig{file=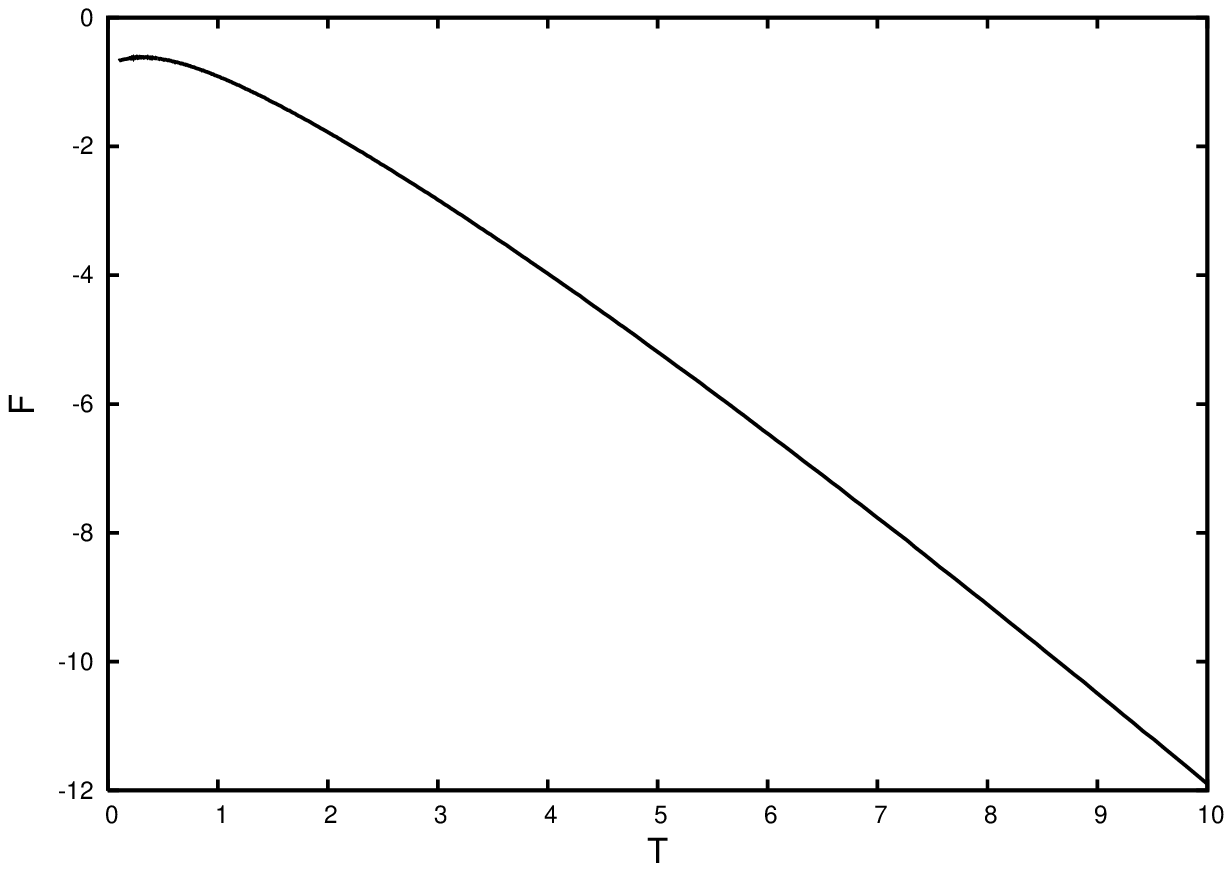,width=0.45\textwidth}\hfill
\epsfig{file=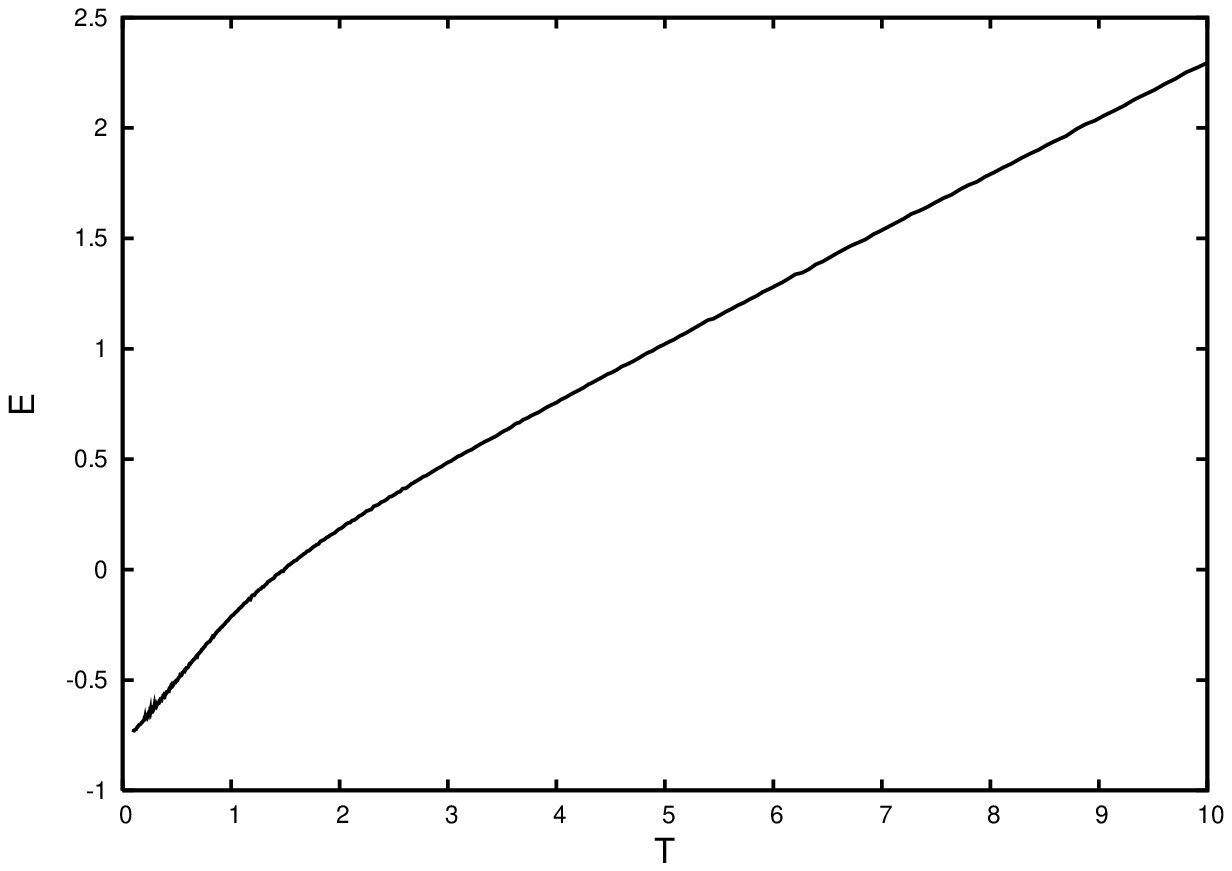,width=0.45\textwidth}\\
\epsfig{file=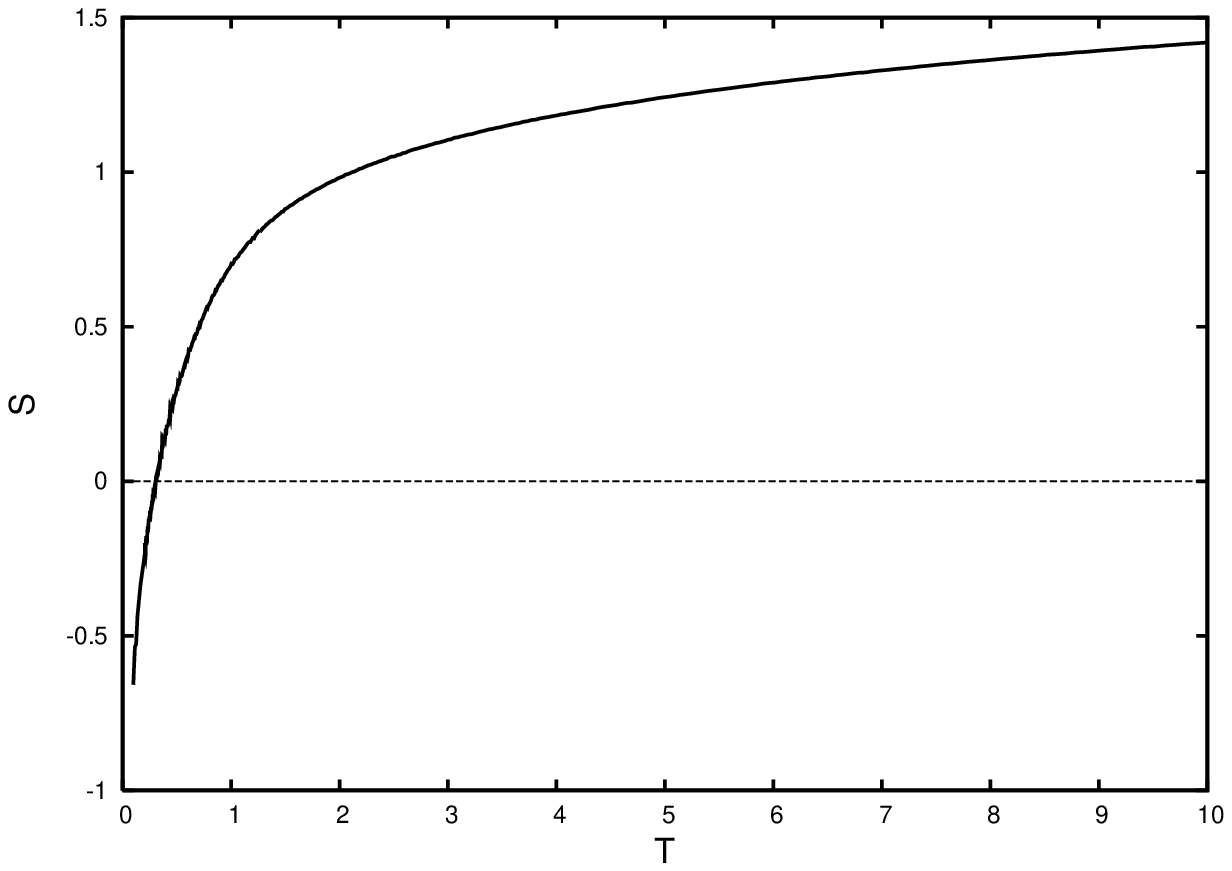,width=0.45\textwidth}\hfill
\epsfig{file=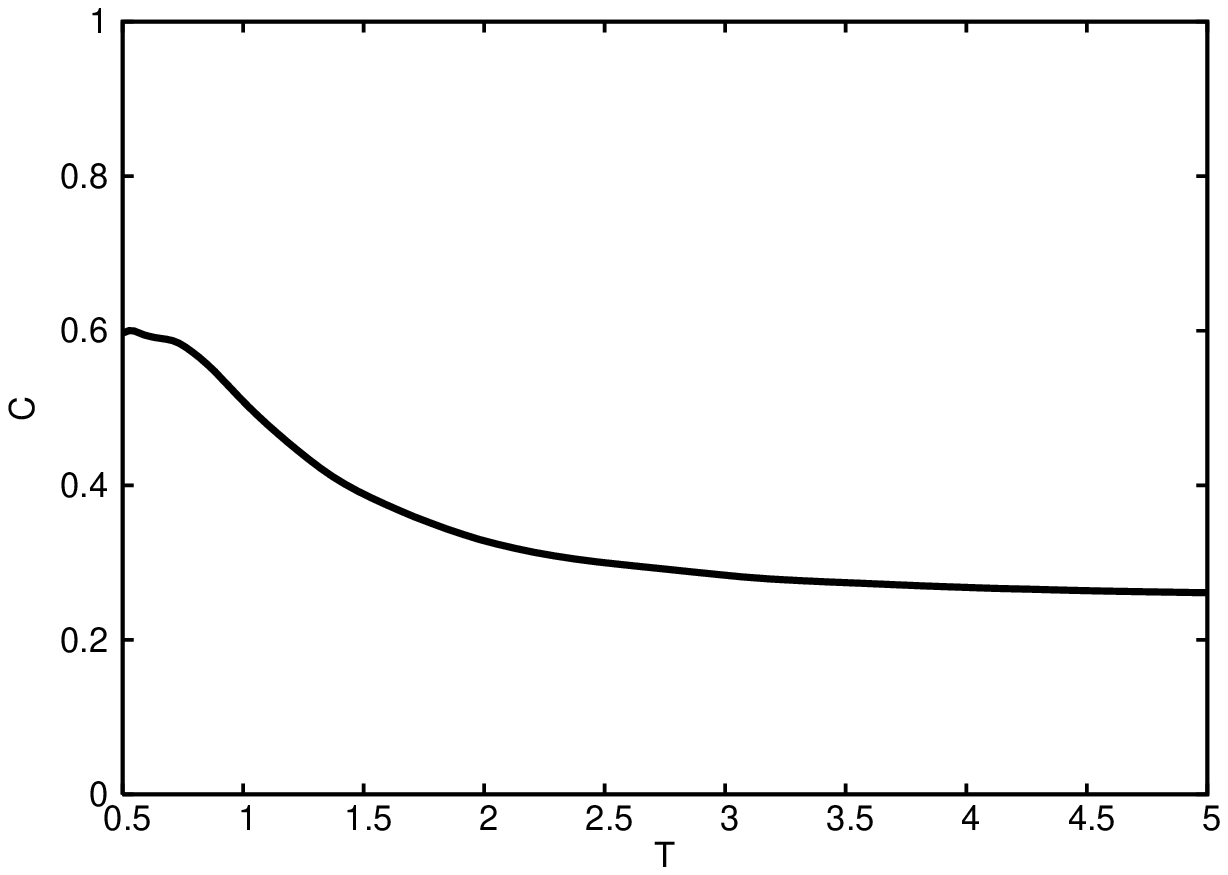,width=0.45\textwidth}
\end{center}
\caption{Thermodynamic functions: free Energy $f$ (first panel), internal energy $E$ 
(second panel), entropy $S$ (third panel), and specific heat $C$, (fourth panel).} 
\label{Fig:thdfunc} 
\end{figure}

In Fig \ref{Fig:thdfunc} we show free energy, internal energy, entropy and specific heat
as functions of temperature, for $\sigma_k=1$ and $g_4=1$, so that $T_c \simeq 0.93$.
The specific heat is  obtained through a numerical differentiation from the internal 
energy, and becomes fairly noisy for temperatures below $T \simeq 0.5$. Notice that the
entropy becomes negative for $T \lesssim 0.31$, i.e. well inside the glassy phase. As 
our system has continous degrees of freedom, no strong conclusions can be drawn from
this observation, however. It must be pointed out, though, that we expect replica symmetry
to be broken throughout the low-temperature phase, and that our results require further
checks, e.g. through simulations. 

Finally, a word on numerics may perhaps be in order. It turns out that a good 
compromise between control and efficiency in evaluating functions like 
$\hat\Psi[\psi,\phi]$ on a sufficiently dense grid of points $\vv$ 
via (\ref{Zpsiphi}) as required for the population dynamics algorithm, is to use
Gaussian quadratures with sufficiently many grid points. Further acceleration of
numerical procedures is possible by evaluating functions initially on coarser 
grids, and obtaining their values on the fine grids required for further processing
via suitable interpolation algorithms. 

\section{Conclusion}						\label{sec:concl}

We have introduced and solved a wide class of random models with finite-connectivity
for the description of amorphous solid phases. The description is in terms of deviations
of particle positions from a set of reference positions. The framework is kept general
in the sense that we do not need not restrict the nature of the pair potential, nor 
the nature of the connectivity distribution, as long as they allow taking a meaningful
thermodynamic limit. By varying the connectivity distributions or the nature of 
interaction potentials we expect to be able to describe, at least in a schematic 
manner, a fairly wide spectrum of different physical systems.

In the present paper we have looked in detail only at harmonically coupled systems
with scalar degrees of freedom, thought to provide an appropriate description of
systems in their gel-phase, and at a model with local quartic anharmonicities
which could be used to describe glassy low-temperature properties. The transition
to the phase with frozen glassy order in the model considered here is continuous,
as in the corresponding fully connected variant. In this respect, the description of
glass-transition physics within this model must still be regarded as deficient. This
may or may not be due to the fact that the low-order expansions of the interaction
energy chosen in the present paper can be considered as adequate description of the 
physics only in the limit of low temperatures.

One promising possibility to improve the description of glass-transition physics 
within the present model-class would be to look at models with {\em non-confining\/} 
interaction potentials. This would inter alia allow obtaining more adequate descriptions 
of high temperature phases, as phases with vanishing inverse localization lengths
(delocalized particles). It remains to be seen whether this modification could put
our model into the class for which glass-transition physics would be described by the 
succession of dynamic freezing and entropy-crisis scenario thought to be appropriate
for structural glasses.

Other problems we might be able to shed some analytic light on are correlations 
between coordination and effective potential statistics, or --- by looking at 
mixtures of different particle species  --- the correlations between particle 
species and their enhanced or depressed role in the formation of tunneling centres
in amorphous systems.

Another open question is related to the fact that the present models --- unlike their 
fully connected counterparts --- are not completely described in terms of ensembles 
of effective non-interacting single-site problems. The two-particle contribution 
appearing in the free energy is a consequence of the fact that finite connectivity 
systems of the type considered here would no longer support simple single particle 
excitations. It is conceivable that the two-particle contribution would allow an
interpretation similar to the `heat-bath' of phonons acting on tunnelling systems
that is introduced in an ad-hoc manner into phenomenological models of glassy 
low-temperature anomalies \cite{And+72,Phil72,Jaeckle72}.

We intend to look into some of these questions in future publications.

\paragraph{Acknowledgements} It is a pleasure to thank ACC Coolen,  J Hatchett, I 
Perez-Castillo, and P Sollich for many illuminating discussions and helpful remarks.

\appendix

\section{Details of the Replica Analysis}		\label{app:replica_det}

\subsection{Averaging the Replicated Partition Function}
Given the distribution (\ref{P_ofC}) of connectivity matrices, the average 
(\ref{reppartfunc}) of the replicated partition function has the following 
structure
\beq
\Bra Z_N^n\Ket_{\bC,\phi}
={\cT\ov\cN}\ev{
\Bra \prod_{(i,j)}\Tr_{c_{ij}}~p_C(c_{ij})~\de_{c_{ij},c_{ji}}
\prod_i~\de_{L_i,\sum_{j(\neq i)}c_{ij}}~  Z_N^n\Ket_{\phi}
\ov
\prod_{(i,j)}\Tr_{c_{ij}}~p_C(c_{ij})~\de_{c_{ij},c_{ji}}
\prod_i~\de_{L_i,\sum_{j(\neq i)}c_{ij}}}\ .
\label{CAv}
\eeq
Note that both the $c_{ij}$-summations and the average over local pair potentials 
$\phi_{ij}$, factor w.r.t the pairs $(i,j)$. 

It should be mentioned that other, equivalent ways of introducing the ensemble
of connectivity matrices are possible. E.g. one could include summations over
all possible assignments of local connectivities $\{L_i\}$ compatible with a
given $P_C(L)$ in both $\cT$ and $\cN$ in (\ref{CAv}), without changing final
results, as this would just create contributions coming from the entropy of
the connectivity distribution in $\cT$ and $\cN$, which cancel. Similarly,
one could omit including the single bond distributions $p_C(c_{ij})$ in $\cT$ 
and $\cN$, as long as the constraints required by the desired connectivity 
contribution $P_C(L)$ are kept --- at the cost of considerably complicating the 
algebra, and without changing final results. In what follows, we stick to the
version embodied in (\ref{CAv}).

To evaluate (\ref{CAv}), one uses integral representations of the form 
(\ref{Kron_delt}) to express the constraints on the local coordinations $L_i$.
The normalization constant $\cN$ in (\ref{CAv}) is then transformed into an
integral expression, which can be evaluated via the saddle point method, along 
the following transparent sequence of steps
\bea
\cN  &=     &
\prod_{(i,j)}\Tr_{c_{ij}}~p_C(c_{ij})~\de_{c_{ij},c_{ji}}~
\prod_i~\oint{\d z_i\ov2\pi\i z_i}z_i^{(\sum_{j(\neq i)}c_{ij}-L_i)}
\nn\\&=     &
\prod_i~\oint{\d z_i\ov2\pi\i z_i}z_i^{-L_i}
\prod_{(i,j)}\Tr_{c_{ij}}~p_C(c_{ij})~(z_iz_j)^{c_{ij}}
\nn\\&\simeq&
\prod_i~\oint{\d z_i\ov2\pi\i z_i}z_i^{-L_i}
\e^{\lh{C\ov2N}\sum_{i,j}(z_iz_j-1)\rh}
\nn\\&=     &
\int \frac{\d \hrh_0\d \rho_0}{2\pi/N}~\exp\lh N\lc{C\ov2}(\rho_0^2-1)-\hrh_0\rho_0\rc\rh
\prod_i~\oint{\d z_i\ov2\pi\i z_i}z_i^{-L_i}\e^{\hrh_0z_i}
\nn\\&=     &
\int \frac{\d \hrh_0\d \rho_0}{2\pi/N}~\exp\lh N\lc{C\ov2}(\rho_0^2-1)-\hrh_0\rho_0+
\frac{1}{N}\sum_i ~\ln \left({\hrh_0^{L_i}\ov L_i!}\right)\rc\rh
\label{N}
\eea
Here, we used large $N$ asymptotics to re-exponentiate the $c_{ij}$ averages in line 2, 
and introduced $\rho_0\ev{1\ov N}\sum_iz_i$, enforced by the conjugate $\hrh_0$, 
to obtain an expression that can be evaluated by the saddle point method, and finally
exploited the fact that (\ref{Kron_delt}) entails
\beq
\oint{\d z\ov2\pi\i z}z^{-L}f(z)=
\left.{1\ov L!}{\partial^L f(z)\ov\partial z^L}\right|_{z=0}\ .
\eeq
As the $L_i$ are distributed according to $P_C(L)$, the last contribution in the exponent
in (\ref{N}) can be expressed as the $L$-average $\sum_L P_C(L)~\ln\Big({\hrh_0^L\ov L!}\Big)$
by the law of large numbers. In the saddle point of (\ref{N}), i.e. $(\rho_0=1,\hrh_0=C)$ 
we finally have:
\beq
\cN \sim \exp\lh N\lc\sum_LP_C(L)\ln\lh{C^L\ov L!}\rh-C\rc\rh
\eeq
%
%
Given any typical realization of local potentials $\{V_i,i=1,..,N\}$, and
using the shorthands (\ref{rpot}) for the replicated bond and site weights
we have for nominator $\cT$ in (\ref{CAv}), following similar lines of reasoning:
\bea
\cT&=     &
\prod_i\int \d \tvu_i~\cU_s(\tvu_i,V_i)~
\prod_{(i,j)}\Tr_{c_{ij}}~p_C(c_{ij})~\de_{c_{ij},c_{ji}}~
\Bra\lh{\ov}\cU_b(\tvu_i,\tvu_j,\phi)\rh^{c_{ij}}\Ket_\phi
\prod_i \de_{L_i,\sum_{j(\neq i)}c_{ij}}
\nn\\
   &\simeq&\prod_i\int \d \tvu_i~\cU_s(\tvu_i,V_i)
       \oint{\d z_i\ov2\pi\i z_i}z_i^{-L_i}~~
\e^{{C\ov2N}\sum_{i,j} z_iz_j~\Bra \cU_b(\tvu_i,\tvu_j,\phi){\ov}\Ket_\phi-
{CN\ov2}}
\nn\\
   &=     &\int\cD\rho\cD\hrh~
\exp\lh N\lc{C\ov2}(G_b[\rho]-1)-G_m[\rho,\hrh]\rc\rh
\prod_i\int \d \tvu_i~\cU_s(\tvu_i,V_i)
           \oint{\d z_i\ov2\pi\i z_i}z_i^{-L_i}~
           \e^{z_i\hrh(\tvu_i)}
\nn\\
   &=     &\int\cD\rho\cD\hrh~\exp\lh N\lc
{C\ov2}(G_b[\rho]-1)-G_m[\rho,\hrh]+G_s[\hrh]\Big]\rc\rh
\label{T}
\eea
The definition (\ref{repdens}) of the replica density enforced by a conjugate 
$\hrh(\tvu)$ was used to express the result in terms of a path-integral that
can be evaluated by the saddle point method.

The three functionals appearing in (\ref{T}) are given as
\bea
G_b[\rho     ~]&=&\Bra \int \d \tvu~\d
\tvv~\rho(\tvu)~\cU_b(\tvu,\tvv,\phi)\rho(\tvv)
\Ket_\phi
\nn\\
G_m[\rho,\hrh~]&=&\int \d \tvu~     ~\hrh(\tvu)~                 \rho(\tvu) 
\\
G_s[     \hrh~]&=&\sum_{L}P_C(L)\Bra\ln\int \d \tvu~{\hrh^L(\tvu)\ov L!}
~\cU_s(\tvu,V)\Ket_V
\eea
where the law of large numbers is used to write $G_s$ as an average over the
site disorder (in the $L_i$ and the $V_i$).

\subsection{Replica Symmetry}
We proceed by making an ansatz for $\rho$ and $\hrh$ which amounts to assuming 
unbroken replica symmetry (RS):
\bea
\rho(\tvu) &=&\int\cD\psi~\pi [\psi]~
\prod_a{\exp[-\be\psi(\vu^a)]\ov Z[\psi]}
\\
\hrh(\tvu) &=&\int\cD\hps~\hpi[\hps]~
\prod_a{\exp[-\be\hps(\vu^a)]\ov Z[\hps]}\ ,
\eea
where $\pi$ and $\hpi$ are functionals over the space of single site potentials
and $Z[\psi]$ and $Z[\hps]$ are defined via (\ref{Zf}). We then obtain:
\bea
G_b[\rho     ~]& \simeq &\rho_0^2+n~
\int \cD\pi[\psi_1]\cD\pi[\psi_2]
\Bra\ln\lh{Z_2[\psi_1,\psi_2,\phi]\ov Z[\psi_1]Z[\psi_2]}\rh\Ket_\phi
\nn\\
G_m[\rho,\hrh~]& \simeq &\rho_0\hrh_0+n~
\int \cD\pi[\psi]\cD\hpi[\hps]~\ln\lh{Z[\psi+\hps]\ov Z[\psi] Z[\hps]}\rh
\\
G_s[     \hrh~]& \simeq &\sum_{L}P_C(L)\lh\ln\Big({\hrh_0^L\ov L!}\Big)+
n\int\{\cD\hpi\}_L\Bra\ln\lh{Z[\sum_\ell^L\hps_\ell+V]\ov
\prod_{\ell=1}^L Z[\hps_\ell]}\rh\Ket_V\rh
\nn
\eea
where the definition (\ref{Z2}) of $Z_2$, the convention 
$\{\cD\hpi\}_L\ev \prod_\ell^L\cD\hpi[\hps_\ell]\ $, and the abbreviations
      $\rho_0       \ev \int\cD\pi [\psi]\ $,
and   $\hrh_0       \ev \int\cD\hpi[\hps]\ $ 
for full integrals over the functionals $\pi$ and $\hpi$ have been used.

We first solve the saddle point equations to $\cO(1)= \cO(n^0)$ in $n$, to obtain 
$\rho_0=1$, and $\hrh_0=C$. This entails that the $\cO(n^0)$ contributions to $\cT$
cancel with $\cN$, and we finally obtain
\bea
\Bra Z_N^n\Ket_{\bC,\phi} & \sim & \int\cD\pi~\cD\hpi
\exp\lh nN\lc{C\ov2} \cG_b[\pi]-C\cG_m[\pi,\hpi]+\sum_L P_{C}(L)~\cG_{s,L}[\hpi]
          \rc\rh\ ,
\eea
in which $\pi$ and $\hpi$ are now {\em normalised\/} functionals, and where
\bea
\cG_b[\pi     ]& \simeq & \int\cD\pi[\psi_1]~\cD\pi[\psi_2]~
\Bra\ln\lh{Z_2[\psi_1,\psi_2,\phi]\ov Z[\psi_1]Z[\psi_2]} \rh\Ket_\phi~,
\\
\cG_m[\pi,\hpi]& \simeq & \int\cD\pi[\psi ]~\cD\hpi[\hps ]~
      \ln\lh{Z[\psi+\hat\psi] \ov Z[\psi] Z[\hps]} \rh~,
\\ 
\cG_{s,L}[\hpi]& \simeq & \int\{\cD \hpi\}_L~
\Bra\ln\lh {Z\Big[\sum_{\ell=1}^L \hps_\ell+V\Big]\ov
\prod_{\ell=1}^L Z[\hps_\ell]} \rh\Ket_V~.
\eea
The saddle point equations w.r.t. the normalised $\pi,~\hpi$ can then be
expressed in the form (\ref{SP_hpi}) and (\ref{SP_pi}) as given in Sect 
\ref{sec:rs}.
%
%
%
%
%
\subsection{Breaking Replica Symmetry}
It is expected that replica symmetry will be broken in phases with frozen glassy
order. We will in the present paper not go on to solve our models assuming phases
with broken replica symmetry. However we would here like to present at least 
the {\em ansatz\/} which would describe phases with one level of replica symmetry
breaking in the spirit of Parisi' hierarchical scheme. It is based on grouping
the $n$ replica in $n/m$ blocks containing $m$ replica each, and assuming symmetry
of the solution w.r.t. permutations of replica within blocks though not between
blocks. This would lead to 
\bea
\rho(\tvu) &=&\int\cD\pi~\cP[\pi] \prod_{k=1}^{n/m} \int\cD\psi_k~\pi [\psi_k]~
\prod_{a=(k-1)m +1}^{km} {\exp[-\be\psi_k(\vu^a)]\ov Z[\psi_k]}
\\
\hrh(\tvu) &=&\int\cD\hpi~\hat\cP[\hpi]  \prod_{k=1}^{n/m}  \int\cD\hps_k~\hpi[\hps_k]~
\prod_{a=(k-1)m +1}^{km} {\exp[-\be\hps_k(\vu^a)]\ov Z[\hps_k]}\ ,
\eea
as ansatz for the replica-density (\ref{repdens}) and the conjugate required to
enforce its  definition, respectively. I.e., we get a functional superposition
of products of block-replica densities, each of which a functional superposition
of products of single-replica Gibbs distributions over the replica within a block.

This ansats would again allow to formulate the $n\to 0$-limit of the theory, and
would lead to fixed point equations for the weight functions $\cP$, $\pi$, $\hat\cP$
and $\hat\pi$ that can be cast into a form which would admit a solution in terms
of a stochastic population-based algorithm --- an algorithm, however, which requires
to maintain an {\em ensemble of populations\/} of the type used in the RS version of the 
theory. In addition, there is the stationarity requirement on the Free Energy
with respect to the partitioning parameter $m$. The numerical effort required to
solve this problem will be considerable. 

\section{Free Energy, Thermodynamic Functions}		\label{app:thd_func}

The replica symmetric expression of the free energy as a funcional of
$\pi$ and $\hpi$  obtained from (\ref{Zfinal})-(\ref{cGs}) initially gives
\bea
- \be f(\be) &=& {C\ov 2}\int \cD\pi[\psi]\cD\pi[\psi']
\Bra \ln\lv{Z_2[\psi,\psi',\phi]\ov Z[\psi]Z[\psi']} \rv\Ket_\phi
-C\int \cD\pi[\psi]\cD\hpi[\hat\psi]~\ln\lv{Z[\psi+\hat\psi] 
\ov Z[\psi]\hZ[\hat\psi]}\rv
\nn\\
& & + \sum_L P_{C}(L) \int \{\cD \hpi\}_L
\Bra\ln\lv {Z\Big[\sum_{\ell=1}^L \hat\psi_\ell +V \Big]\ov\prod_{\ell=1}^L 
\hZ[\hat\psi_\ell]} \rv \Ket_V \ ,
\label{fRSinitial}
\eea
in which $\pi$ and $\hat\pi$ are taken to be solutions of the fixed-point 
equations.

Replacing logarithms of products (quotients) by sums (differences) of logaritms,
and recalling  and $\langle L \rangle = C$, we first get
\bea
- \be f(\be) &=& {C\ov 2}\int \cD\pi[\psi]\cD\pi[\psi']
\Bra\ln\lv Z_2[\psi,\psi',\phi] \rv\Ket_\phi
-C\int \cD\pi[\psi]\cD\hpi[\hat\psi]~\ln\lv Z[\psi+\hat\psi]\rv
\nn\\
& &+ \sum_L P_{C}(L) \int \{\cD \hpi\}_L~
\Bra\ln\lv Z\Big[\sum_{\ell=1}^L \hat\psi_\ell + V\Big]\rv \Ket_V\ .
\label{RSFreeEn1}
\eea
In a second step, we express $\hpi[\hat\psi]$ appearing in the second integral
by an integral over $\pi[\psi]$ via (\ref{SP_hpi}) to get
\bea
\int \cD\pi[\psi]\cD\hpi[\hat\psi] \ln\lv Z[\psi+\hat\psi]\rv 
&=&\int \cD\pi[\psi]\cD\pi[\psi']~
   \int \cD\hat\psi \Bra \de\lv \hat\psi-\hat\Psi[\psi',\phi]\rv\Ket_\phi \ln\lv 
   Z[\psi+\hat\psi]\rv
\nn\\
&=&\int \cD\pi[\psi]\cD\pi[\psi'] \Bra \ln \lv
   Z[\psi+\hat\Psi[\psi',\phi]]\rv \Ket_\phi
\nn\\ 
&=&\int \cD\pi[\psi]\cD\pi[\psi'] \Bra\ln\lv Z_2[\psi,\psi',\phi] \rv\Ket_\phi\ .
\nn
\eea
In the last step, the definition (\ref{PsiH}), (\ref{Zpsiphi}) of 
$\hat\Psi[\psi',\phi])$ was invoked. We thus note that the first and second 
contribution to (\ref{RSFreeEn1}) can be combined so as to finally obtain 
(\ref{RSFreeEn}). 

The internal energy is obtained from the free energy via the relation $E(\be)= 
{\partial \ov \partial \be} (\be f(\be))$.

\section{Bethe-Peierls Method}				\label{app:BPMethod}
The replica-symmetric fixed point equations can be derived as resulting from
a recursive evaluation of the partition function of the system described by 
an energy function of the form
\beq 
U_{\rm int}=\sum_{(i,j)} c_{ij}\phi_{ij}(u_i-u_j)  + \sum_i V(u_i)
\eeq 
assuming that the collection of nodes form a tree-like structure. Scalar degrees
of freedom and site-independent on-site potentials $V$ are assumed for simplicity.

The restricted partition function computed at fixed value $u_i$ for the  
coordinate at $i$ is
\beq
Z_i(u_i) = \exp\{-\be V(u_i)\} \int \prod_{j\in \cN(i)} \rd u_j \exp\{-\be
\phi_{ij}(u_i-u_j)\} Z_{j|i}(u_j)
\eeq
in which $\cN(i)$ denotes the set of vertices connected to vertex $i$, the
$Z_{j|i}(u_j)$ are the restricted partition functions of the subtrees rooted in
$j$ -- exluding the tree rooted in node $i$ -- evaluated at the value $u_j$ of
the coordinate at the root-node $j$. If the system were tree-like, the $u_j$
integrals in this expression would factor, 
\beq
Z_i(u_i) = \exp\{-\be V(u_i)\} \prod_{j\in \cN(i)}
\int \rd u_j \exp\{-\be \phi_{ij}(u_i-u_j)\} 
Z_{j|i}(u_j),
\label{factoring}
\eeq
and we would have the recursion
\beq
Z_{j|i}(u_j) =  \exp\{-\be V(u_j)\} \prod_{k\in \cN(j)\setminus i}
\int \rd u_k \exp\{-\be \phi_{jk}(u_j-u_k)\} 
Z_{k|j}(u_k)\ .
\eeq
This recursion is the basis for the iterative Bethe-Peierls method for solving
statistical mechanical problems on tree-like structures.

The system we are considering is, however, only locally tree-like, i.e. there
are loops with lengths of $\cO(\ln N)$. Factorization is, therfore, in principle only
approximate. However, if the system has only a single thermodynamic state and
correlations decay exponentially with distance (this is the situation of
unbroken replica symmetry), then factorization would be asymptotically exact in
the thermodynamic limit, as common  ancestors of two sites neighbouring on a
given site occur typically at distance $\cO(\ln N)$ if the given site is
removed, entailing that correlations are negligible in the thermodynamic limit.

One introduces cavity potentials $\hat \psi_{j|i} (u_i)$ associated with
the factors appearing in (\ref{factoring}) via
\beq
\exp\{-\be \hat \psi_{j|i} (u_i)\} = \int \rd u_j \exp\{-\be
\phi_{ij}(u_i-u_j)\}  Z_{j|i}(u_j)
\label{hati0}
\eeq
For the restricted partition functions $Z_{j|i}(u_j)$  of the sub-trees
themselves we have
\beq
Z_{j|i}(u_j) = \exp\{-\be V(u_j)\}  \prod_{k\in \cN(j)\setminus i} \exp\{-\be
\hps_{j|i} (u_i) \}
\eeq
In terms of the cavity potentials $\hps_{j|i}$ and the free energies
$\psi_{j|i}(u_j)$ associated with restricted sub-tree partition functions via
\beq
Z_{j|i}(u_j)=\exp\{-\be \psi_{j|i}(u_j)\}
\eeq
one has the recursion
\bea
\psi_{j|i}(u_j) &=& V(u_j) + \sum_{k\in \cN(j)\setminus i} \hat \psi_{k|j}(u_j) 
\label{itpsi}\\
\hps_{k|j}(u_j) &=&  -\be^{-1} \ln \int \rd u_k \exp\{-\be
\phi_{jk}(u_j-u_k) -\be \psi_{k|j}(u_j)\}\nn \\
	&\ev& -\be^{-1} \ln Z_{\psi_{k|j}\phi_{jk}}
\label{itpsih}
\eea
The replica symmetric fixed-point equations derived in Sect. (\ref{sec:rs})
are now recovered by observing that due to the locally varying coordination and
the heterogeneity of the interaction potentials $\phi_{ij}$ associated
with the edges $(i,j)$ of the tree-like graph, the cavity potentials and the
restricted sub-tree free energies are random. Hence the iteration has to be
formulated in a probabilistic setting as an iteration for probabilities of
finding certain restricted sub-tree free energies $\psi$ and cavity potentials
$\hat\psi$. For a homogeneously disordered (locally) tree-like graph this would
lead to self-consistency conditions that have to hold in the thermodynamic
limit. Using Eqs. (\ref{itpsi}) and (\ref{itpsih}) one finds
\bea
    \pi[\psi] &=& \sum_{L > 0}{L\ov C} P_C(L) \int \prod_{k=1}^{L-1}
                  \cD\hat\pi[\hps_k] \delta \Big[\psi - \Psi_{L-1}[\{\hps_k\}]\Big]
\\ 
\hat\pi[\hps] &=& \int \cD \pi[\hps] \Bra\delta[\hps -
                  \hat\Psi[\psi,\phi]]\Ket_\phi
\eea
for the self-consistency conditions, on a graph with connectivity distribution
$P_C(L)$ of average coordination $C$, and 
$\Psi_{L-1}$ and $\hat\Psi[\psi,\phi_\om]$ are as defined via
(\ref{Psi}) and (\ref{PsiH}).

\bibliography{../../MyBib}
\end{document}